%% file: mrs977.tex
\documentstyle[12pt,graphics,epsf,cites]{article}

\input amssym.def
\input amssym

\begin{document}
\numcite
\def\onlinecite#1{[\citenum{#1}]}

\baselineskip+0.6cm 

\begin{center}
{\bf {THIRD-GENERATION TB-LMTO}}

\smallskip
\end{center}

\medskip \noindent
O.K. ANDERSEN, C. ARCANGELI, R.W. TANK,

\noindent
T. SAHA-DASGUPTA, G. KRIER, O. JEPSEN, and I. DASGUPTA.

\noindent
Max-Planck Institut f\"{u}r Festk\"{o}rperforschung, Stuttgart, Germany

\smallskip

\bigskip

\noindent
ABSTRACT

\smallskip

\smallskip We describe the screened Korringa-Kohn-Rostoker (KKR) method and
the third-generation linear muffin-tin orbital (LMTO) method for solving the
single-particle Schr\"{o}dinger equation for a MT potential. In the screened
KKR method, the eigenvectors $c_{RL,i}$ are given as the non-zero solutions,
and the energies $\varepsilon _i$ as those for which such solutions can be
found, of the linear homogeneous equations: $\sum_{RL}K_{R^{\prime
}L^{\prime },RL}^a\left( \varepsilon _i\right) \,c_{RL,i}=0,$ where $%
K^a\left( \varepsilon \right) $ is the screened KKR matrix. The screening is
specified by the boundary condition that, when a screened spherical wave $%
\psi _{RL}^a\left( \varepsilon ,{\bf r}_R\right) $ is expanded in spherical
harmonics $Y_{R^{\prime }L^{\prime }}\left( {\bf \hat{r}}_{R^{\prime
}}\right) $ about its neighboring sites ${\bf R}^{\prime },$ then each
component either vanishes at a radius, $r_{R^{\prime }}{\rm =}a_{R^{\prime
}L^{\prime }},$ or is a regular solution at that site. When the
corresponding ''hard'' spheres are chosen to be nearly touching, then the
KKR matrix is usually short ranged and its energy dependence smooth over a
range of order 1 Ry around the centre of the valence band. The KKR matrix, $%
K\left( \varepsilon _\nu \right) ,$ at a fixed, arbitrary energy turns out
to be the negative of the Hamiltonian, and its first energy derivative, $%
\dot{K}\left( \varepsilon _\nu \right) ,$ to be the overlap matrix in a
basis of kinked partial waves, $\Phi _{RL}\left( \varepsilon _\nu ,{\bf r}%
_R\right) ,$ each of which is a partial wave inside the MT-sphere, tailed
with a screened spherical wave in the interstitial, or taking the other
point of view, a screened spherical wave in the interstitial, augmented by a
partial wave inside the sphere. When of short range, $K\left( \varepsilon
\right) $ has the two-centre tight-binding (TB) form and can be generated in
real space, simply by inversion of a positive definite matrix for a cluster.
The LMTOs, $\chi _{RL}\left( \varepsilon _\nu \right) ,$ are smooth orbitals
constructed from $\Phi _{RL}\left( \varepsilon _\nu ,{\bf r}_R\right) $ and $%
\dot{\Phi}_{RL}\left( \varepsilon _\nu ,{\bf r}_R\right) ,$ and the
Hamiltonian and overlap matrices in the basis of LMTOs are expressed solely
in terms of $K\left( \varepsilon _\nu \right) $ and its first {\em three}
energy derivatives. The errors of the single-particle energies $\varepsilon
_i$ obtained from the Hamiltonian and overlap matrices in the $\Phi \left(
\varepsilon _\nu \right) $- and $\chi \left( \varepsilon _\nu \right) $
bases are respectively of second and fourth order in $\varepsilon
_i-\varepsilon _\nu .$ Third-generation LMTO sets give wave functions which
are correct to order $\varepsilon _i-\varepsilon _\nu ,$ not only inside the
MT spheres, but also in the interstitial region. As a consequence, the
simple and popular formalism which previously resulted from the
atomic-spheres approximation (ASA) now holds in general, that is, it
includes downfolding and the combined correction. Downfolding to
few-orbital, possibly short-ranged, low-energy, and possibly orthonormal
Hamiltonians now works exceedingly well, as is demonstrated for a
high-temperature superconductor. First-principles $sp^3$ and $sp^3d^5$ TB
Hamiltonians for the valence and lowest conduction bands of silicon are
derived. Finally, we prove that the new method treats overlap of the
potential wells correctly to leading order and we demonstrate how this can
be exploited to get rid of the empty spheres in the diamond structure.

\bigskip

\noindent
INTRODUCTION

\smallskip

There is a need for an {\em intelligible} and {\em accurate}
first-principles electronic-structure method. Our efforts have been directed
towards developing a single-particle basis which, for the application at
hand, can be adjusted to a useful compromise between being {\em short
ranged, minimal,} and {\em accurate.}

\medskip

\noindent \underline{Recent developments of Multiple Scattering Theory}

\smallskip

Since atoms are nearly round, it seems most natural to start out using {\em %
spherical waves,} $j_l\left( \kappa r\right) Y_L\left( {\bf \hat{r}}\right) $
and $n_l\left( \kappa r\right) Y_L\left( {\bf \hat{r}}\right) ,$ where $L%
{\rm =}lm,$ as done when solving Schr\"{o}dinger's equation with the
classical multiple-scattering method due to Korringa, Kohn, and Rostoker
(KKR).\cite{KKR} In this method the scattering by the atom at site ${\bf R}$
is specified by the {\em phase shifts,} $\eta _{Rl}\left( \kappa \right) ,$
of its potential well, and the structure of the solid is specified by a
Hermitian matrix with elements $B_{RL,R^{\prime }L^{\prime }}\left( \kappa
\right) ,$ the {\em structure constants.} In terms of these, the
wave-function coefficients, $c_{RL,i},$ are the solutions of the
homogeneous, linear equations, one for each $R^{\prime }L^{\prime }:$%
\begin{equation}
\sum_{RL}\left[ B_{R^{\prime }L^{\prime },RL}\left( \kappa \right) +\kappa
\cot \eta _{Rl}\left( \kappa \right) \delta _{R^{\prime }L^{\prime
},RL}\right] \,c_{RL,i}=0,  \label{uKKR}
\end{equation}
and the energies, $\varepsilon _i,$ are the values of $\kappa ^2$ $\left(
\equiv \varepsilon \right) $ for which solutions can be found, {\it i.e.}
the determinant of $B\left( \kappa \right) +\kappa \cot \eta \left( \kappa
\right) $ vanishes. There are merely 4--16 equations per atom because all
phase shifts with $l>$ 1--3 vanish. The KKR equations provide the {\em exact 
}solutions of Schr\"{o}dinger's equation, but only for a muffin-tin (MT)
potential, $V\left( {\bf r}\right) \equiv \sum_Rv_R\left( \left| {\bf r-R}%
\right| \right) ,$ which is a superposition of spherically symmetric,
non-overlapping potential wells, $v_R\left( r\right) ,$ of ranges $s_R.$

There are three problems with the KKR method: First of all, a
non-overlapping MT potential is a poor representation of the self-consistent
potential in any, except the most close packed solid. Secondly, the
structure constants have long range, and thirdly, strong energy dependence.
Specifically, the energy dependence of the KKR matrix, $B\left( \kappa
\right) +\kappa \cot \eta \left( \kappa \right) ,$ is not linear like that
of the secular matrix, $H-\varepsilon O,$ for an energy eigenvalue problem
in an energy-independent, possibly non-orthonormal representation. The main
reason for the second and third drawbacks is that the spherical Bessel $%
\left( j_l\right) $ and Neumann $\left( n_l\right) $ functions have long
range and depend on energy. This leads to interferences, which cause long
range and strong energy dependence of the structure matrix. For a crystal
with lattice translations ${\bf T,}$ the Bloch-summed structure matrix, $%
B_{R^{\prime }L^{\prime },RL}\left( \kappa ,{\bf k}\right) \equiv \sum_T\exp
\left( i{\bf k\cdot T}\right) \,B_{\left( R^{\prime }+T\right) L^{\prime
},RL}\left( \kappa \right) ,$ must be evaluated by the Ewald procedure, has
poles at the free-electron parabola, $\kappa ^2=\sum_G\left| {\bf k+G}%
\right| ^2,$ and a branch cut at the bottom of the continuum, $\kappa $=0.

It was recently shown,\cite{Boston} and we shall present a slightly
different proof below, that even when the potential wells {\em overlap,} the
KKR equations do hold to first order in the potential overlap. This, as we
shall demonstrate, allows the use of MT spheres with up to at least 50 per
cent radial overlap [$s_R+s_{R^{\prime }}\lesssim 1.5\left| {\bf R-R}%
^{\prime }\right| $ for $R$ and $R^{\prime }$ denoting nearest neighbors],
and hence treat the potential between the atoms in a more realistic way.
With such large overlaps, the zero of the potential moves from the potential
threshold between the atoms towards the vacuum level, and this means that
the energies for the occupied states are usually negative.

It was furthermore shown,\cite{Boston} that transformation to {\em linear
combinations} of spherical Bessel and Neumann functions, so-called {\em %
screened} spherical waves, characterized by a set of background phase shifts 
$\alpha _{Rl}\left( \kappa \right) ,$ can remove the long range {\em and}
the strong energy dependence from the structure matrix, provided that the
energy is not too high. This screening transformation may be expressed as: 
\begin{equation}
\left| n^\alpha \right\rangle =\left| n\right\rangle \left( 1-\frac{\tan
\alpha }\kappa \,B^\alpha \right) ,\;{\rm where}\;\tan \eta ^\alpha =\tan
\eta -\tan \alpha ,\;{\rm and}\;\left[ B^\alpha \right] ^{-1}=B^{-1}+\frac{%
\tan \alpha }\kappa .  \label{trf1}
\end{equation}
These are, respectively, vector-, scalar-, and matrix equations. The
superscript labels the representation. In the first equation, we have used a
notation in which $\left| n\right\rangle $ is a row vector of functions with
components $n_l\left( \kappa r_R\right) Y_{lm}\left( {\bf \hat{r}}_R\right) $
and where ${\bf r}_R\equiv {\bf r-R.}$ The last equation involves inversion
of the matrix $B+\kappa \cot \alpha .$ For a set of background phase shifts,
which are known to give short range, this inversion may be performed in real
space and the screened KKR method is basically a first principles
tight-binding (TB) method. At the time,\cite{Boston} however, the relation
between range and background phase shifts was poorly understood. This
problem was solved later\cite{Trieste} by expressing the background phase
shifts in terms of their hard-sphere radii, $a_{RL},$ defined by 
\begin{equation}
\tan \alpha _{RL}\left( \kappa \right) \equiv j_l\left( \kappa a_{RL}\right)
/n_l\left( \kappa a_{RL}\right) ,  \label{alpha}
\end{equation}
or equivalently, by letting the background phase shifts be those of
repulsive potential wells.\cite{Zeller} Looked upon in this way, the role of
the confinement is to push the bottom of the continuum up in energy with
respect to the floor of the MT potential, and thereby, to leave below a
range of energies in which the confined wave-equation solutions are
localized and which, in order to be useful, should include the range of the
occupied bands for the real (attractive) potential. With the definition (\ref
{trf1}), the screened spherical waves, $\left| n^\alpha \right\rangle ,$ are
still quite energy dependent, but only due to their normalization. A more
suitable normalization followed naturally from the hard-sphere point of view.%
\cite{Trieste}

Finally, it was pointed out\cite{Boston} that the screening transformation
may also be used to remove unwanted channels from the KKR equations by
choosing, for those channels, the background phase shifts equal to the real
phase shifts, $\alpha \left( \kappa \right) =\eta \left( \kappa \right) $.
This is a transformation to a {\em minimal basis.}

With these three recent developments, we have the basic ingredients for the
Schr\"{o}dinger part of an {\em intelligible and accurate} method, the
third-generation LMTO method.

\medskip

\noindent \underline{Earlier developments; the Atomic Spheres Approximation}

\smallskip

\smallskip \ The attempts to develop from the KKR method an {\em intelligible%
} first-principles method were initiated 25 years ago\cite{An} and
overlapping spheres, the two-centre interpretation, and screening
transformations\cite{An84} have been used routinely for a long time. The new
development,\cite{Trieste} which started six years ago,\cite{Boston} and
which will be further elaborated on in the present paper, aims at making the
method also {\em accurate}\ without loss of simplicity and elegance. Whereas
in the earlier developments reduced range and energy dependence were
achieved through a physically motivated approximation, namely the
atomic-spheres approximation (ASA), this is now achieved {\em exactly,} and
that in turn, allows a {\em controlled} approximation for the potential
overlap.

The ASA\cite{An,Skriver} consists of letting the MT spheres overlap to the
extent that they become space filling, whereby the interstitial region is
effectively eliminated so that one may neglect the energy dependence of the
wave functions in this region and, hence, the energy dependence of the
structure matrix. The remaining energy dependence now occurs only along the
diagonal of the KKR-ASA matrix where it enters through the radial
logarithmic derivatives evaluated at the atomic sphere, 
\begin{equation}
D\left\{ \phi _{Rl}\left( \varepsilon ,s_R\right) \right\} \equiv s_R\phi
_{Rl}^{\prime }\left( \varepsilon ,s_R\right) /\phi _{Rl}\left( \varepsilon
,s_R\right) .  \label{LogDer}
\end{equation}
If, as is usually done, the kinetic energy in the interstitial is taken to
be zero $\left( \kappa ^2=0\right) $, the suitably renormalized
KKR-ASA equations become: 
\[
\sum_{RL}\left[ {\sf S}_{R^{\prime }L^{\prime },RL}^0-{\sf P}_{Rl}^0\left(
\varepsilon \right) \delta _{R^{\prime }L^{\prime },RL}\right]
\,c_{RL,i}=0,\quad {\rm where} 
\]
\[
{\sf P}_{Rl}^0\left( \varepsilon \right) \equiv 2\left( 2l+1\right) \left(
\frac w{s_R}\right) ^{2l+1}\frac{D\left\{ \phi _{Rl}\left( \varepsilon
,s_R\right) \right\} +l+1}{D\left\{ \phi _{Rl}\left( \varepsilon ,s_R\right)
\right\} -l}\approx \left[ \frac{\Delta _{Rl}}{\varepsilon -C_{Rl}}+\gamma
_{Rl}\right] ^{-1} 
\]
are the potential functions for well $v_R\left( r\right) ,$ and $C_{Rl},$ $%
\Delta _{Rl},$ and $\gamma _{Rl}$ are potential parameters. ${\sf S}^0$ is
the structure matrix given by: 
\begin{equation}
{\sf S}_{{\rm on\,site}}^0=0,\;{\sf S}_{ss\sigma }^0=-2\left( w/d\right) ,\;%
{\sf S}_{sp\sigma }^0=2\sqrt{3}\left( w/d\right) ^2,\;,\;{\sf S}_{dd\left(
\sigma ,\pi ,\delta \right) }^0=10\left( w/d\right) ^5\left( -6,4,-1\right) ,
\label{two}
\end{equation}
when we choose ${\bf R}^{\prime }-{\bf R\equiv \hat{z}}d$ and $w$ is an
arbitrary length scale, usually chosen to be the average Wigner-Seitz
radius. The structure matrix thus consists of effective hopping integrals.
For monatomic crystals, this gave rise to the concept of canonical bands.%
\cite{An} However, the $d^{-l-l^{\prime }-1}$-decay of the hopping integral
between orbitals of angular-momentum characters $l$ and $l^{\prime }$ is too
slow for a tight-binding scheme, except for $d$- and $f$-orbitals.

It was therefore a breakthrough when it became understood that similarity
transformations could be performed on the KKR-ASA $\left( \kappa ^2=0%
\right) $ equations and could lead to short range.\cite{An84} The
transformation from the bare to a screened representation, specified by
screening constants {\sf a}$_{RL},$ is given by: 
\begin{equation}
{\sf P}^{{\sf a}}\left( \varepsilon \right) ^{-1}={\sf P}^0\left(
\varepsilon \right) ^{-1}-{\sf a}\quad {\rm and}\quad \left[ {\sf S}^{{\sf a}%
}\right] ^{-1}=\left[ {\sf S}^0\right] ^{-1}-{\sf a},  \label{ASAscr}
\end{equation}
which are respectively scalar- and matrix equations. {\sf a} is a diagonal
matrix with elements {\sf a}$_{Rl}.$ The corresponding transformation for
the resolvent, useful for Green-function and CPA calculations,\cite{Green}
is: 
\begin{equation}
\left[ {\sf P}^{{\sf b}}\left( z\right) -{\sf S}^{{\sf b}}\right]
^{-1}=\left( {\sf b}-{\sf a}\right) \frac{{\sf P}^{{\sf a}}\left( z\right) }{%
{\sf P}^{{\sf b}}\left( z\right) }+\frac{{\sf P}^{{\sf a}}\left( z\right) }{%
{\sf P}^{{\sf b}}\left( z\right) }\left[ {\sf P}^{{\sf a}}\left( z\right) -%
{\sf S}^{{\sf a}}\right] ^{-1}\frac{{\sf P}^{{\sf a}}\left( z\right) }{{\sf P%
}^{{\sf b}}\left( z\right) }.  \label{scale1}
\end{equation}
This involves no matrix multiplications, but merely rescaling of matrix
elements. As for TB theory, the screening transformation gave a formalism
for the ''kinetic'' part of the often observed dependence of the hopping
integrals and on-site elements on the environment.\cite{Kanpur} The
screening constants yielding short range were found empirically. The
potential-dependent choice {\sf a}$_{Rl}{\rm =}\gamma _{Rl}$, on the other
hand, makes the energy dependence of the KKR-ASA matrix linear (to second
order) so that $C+\sqrt{\Delta }{\sf S}^\gamma \sqrt{\Delta }\equiv h^\gamma 
$ becomes the Hamiltonian in an orthonormal, but not necessarily
short-ranged basis. The transformation finally made it possible to remove
channels from the KKR-ASA equations by choosing {\sf a}$_{Rl}\left(
\varepsilon \right) ={\sf P}_{Rl}^0\left( \varepsilon \right) ^{-1}$ for
such channels.\cite{Lam} This removal, or ''downfolding'', however,
reintroduced energy dependence of the structure matrix.

When performing density-functional calculations one needs to solve not only
Schr\"{o}dinger's but also Poisson's equation, and with the ASA method this
involves approximating not only the potential but also the charge density by
a superposition of slightly overlapping, spherically symmetric
contributions. This gives a very simple scheme which fails badly in
describing total-energy changes caused by symmetry-lowering distortions,\cite
{Pawlowska} however, {\it e.g.} the ASA can be used for calculation of
pressure-volume relations,\cite{Pettifor} but not for calculation of phonon
frequencies. Moreover, since the potential spheres are supposed to be space
filling in the ASA, open structures can only be treated if the interstices
between the atoms are filled with ''empty'' spheres and this works well only
for structures such as the diamond structure, where the interstices have
high symmetry. Even in such a case, for the description to be intelligible
all empty-sphere channels must be downfolded, and that introduces a rather
strong, non-linear energy dependence of the structure matrix which cannot be
treated in the LMTO-ASA approach to be discussed below.\cite{Lam}

\medskip

\noindent \underline{Linear Muffin-Tin Orbitals of the first and second
generations}

\smallskip

In practice one does not solve the KKR equations, but one uses Green
functions in a short-ranged representation and at complex energies,\cite
{Green} or one solves energy eigenvalue equations, $\sum_{RL}\left[
H_{R^{\prime }L^{\prime },RL}-\varepsilon O_{R^{\prime }L^{\prime
},RL}\right] c_{RL,i}=0,$ which are equivalent with the KKR equations in a
certain energy range around some chosen energy, $\varepsilon _\nu .$ In the
linear muffin-tin orbital (LMTO) method,\cite{Wolley,An} such an eigenvalue
problem is arrived at by using the Raleigh-Ritz variational principle for
the Hamiltonian in a basis of LMTO's constructed from the radial Schr\"{o}%
dinger-equation solutions, $\phi _{Rl}\left( \varepsilon ,r\right) ,$ for
the potential wells and their first energy derivatives, $\dot{\phi}%
_{Rl}\left( \varepsilon ,r\right) ,$ {\em at} the chosen energy, $%
\varepsilon {\rm =}\varepsilon _\nu $. In the interstitial region, the first
and second generation LMTOs use the spherical waves at $\kappa _\nu ^2,$ but
not their first energy derivatives, so that the energy dependence in the
interstitial is suppressed. The second-generation LMTO formalism\cite{An84}
is elegant, but only in the ASA and only if no channels have been
downfolded. Under these conditions, the Hamiltonian and overlap matrices are
expressed solely in terms of the structure matrix and the potential
functions: The structure matrix enters the formalism in the form of a
first-order, two-centre TB-Hamiltonian: $h\equiv {\sf \dot{P}}^{-1/2}\left( 
{\sf S}-{\sf P}\right) {\sf \dot{P}}^{-1/2},$ where as usual ${\sf P}\left(
\varepsilon \right) $ is a diagonal matrix. Here and in the following, the
common superscript {\sf a} is dropped and an omitted energy argument means
that the energy is set to $\varepsilon _\nu .$ In terms of this two-centre
Hamiltonian the LMTO set may be expressed as $\left| \chi \right\rangle
\equiv \left| \phi \right\rangle +\left| \dot{\phi}\right\rangle h,$ a form
which may be regarded as the matrix equivalent of the linear approximation $%
\phi _{Rl}\left( \varepsilon ,r\right) \approx \phi _{Rl}\left( r\right) +%
\dot{\phi}_{Rl}\left( r\right) \left( \varepsilon -\varepsilon _\nu \right) .
$ In the basis of these LMTOs the Hamiltonian and overlap matrices are
respectively: 
\begin{equation}
\left\langle \chi \left| -\Delta +V-\varepsilon _\nu \right| \chi
\right\rangle =\,h\left( 1+oh\right) \quad {\rm and\quad }\left\langle \chi
\mid \chi \right\rangle =\,\left( 1+ho\right) \left( 1+oh\right) +hph,
\label{H1}
\end{equation}
where 
\begin{equation}
o\equiv \,\left\langle \phi \mid \dot{\phi}\right\rangle =\,\frac 1{2!}{\sf 
\ddot{P}}/{\sf \dot{P}}\quad {\rm and\quad }p+o^2\equiv \,\left\langle \dot{%
\phi}^2\right\rangle =\,\frac 1{3!}\stackrel{...}{\sf P}/{\sf \dot{P}}
\label{op1}
\end{equation}
are diagonal matrices and it has been assumed that $\phi _{Rl}\left(
r_R\right) Y_{lm}\left( {\bf \hat{r}}_R\right) $ is normalized to unity in
its sphere, {\it i.e.} that $\left\langle \phi ^2\right\rangle =1.$ The
overlap matrix is seen to be nearly factorized and one may therefore
transform to the L\"{o}wdin-orthonormalized representation, $\left| \chi
^{\bot }\right\rangle \equiv \left| \chi \right\rangle \left\langle \chi
\mid \chi \right\rangle ^{-1/2}\sim \left| \chi \right\rangle \left(
1+oh\right) ^{-1}=\left| \chi ^\gamma \right\rangle ,$ in which one finds
the following expansion for the Hamiltonian: 
\begin{equation}
\left\langle \chi ^{\bot }\left| -\Delta +V-\varepsilon _\nu \right| \chi
^{\bot }\right\rangle =h-hoh+h\left[ oho-\left( ph+hp\right) /2\right]
h+...\;.  \label{H2}
\end{equation}
When {\sf a} gives short range this is a power series in a TB Hamiltonian, $%
h^{{\sf a}}.$ Truncation of this series after the first term yields a
spectrum which is accurate in an energy window of size $\sim \left(
10o\right) ^{-1}=\frac 15{\sf \dot{P}}/{\sf \ddot{P}}$ around $\varepsilon
_v,$ but distorted further away. Adding terms, increases the size of this
window at the expense of including further hoppings. The form (\ref{H2}) has
been useful in recursion calculations\cite{Haydock} for structurally
disordered condensed matter.\cite{Nowak}

For a case like the diamond structure, where one only wants LMTOs centered
on atoms, downfolding of the empty-sphere LMTOs is achieved by
transformation of the structure matrix using: {\sf a}$_E=$ ${\sf P}%
_E^0\left( \varepsilon _\nu \right) ^{-1},$ with $E$ referring to the
empty-sphere channels. The energy is here set to $\varepsilon _\nu $ because
in the LMTO-ASA formalism the structure matrix must be energy independent.
Now, an atom-centered LMTO has a tail which extends into the empty spheres,
and here, it is substituted by the corresponding partial waves. The
atom-centered LMTO is therefore: $\left| \chi _A\right\rangle \equiv \left|
\phi _A\right\rangle +\left| \dot{\phi}_A\right\rangle h_{AA}+\left| \phi
_E\right\rangle \dot{h}_{EA},$ with $\dot{h}_{EA}\equiv \left[ -\partial 
{\sf P}_E^0\left( \varepsilon \right) ^{-1}\left/ \partial \varepsilon
\right| _{\varepsilon _\nu }\right] ^{-1/2}{\sf S}_{EA}\left( \varepsilon
_\nu \right) {\sf \dot{P}}_A^{-1/2}.$ This is the way in which the energy
dependence of {\sf a}$_E\left( \varepsilon \right) $ enters, but only to
linear order. The overlap matrix $\left\langle \chi _A\mid \chi
_A\right\rangle $ will now contain the term $\dot{h}_{AE}\dot{h}_{EA}$
involving $A-E-A$ hoppings, in addition to the terms in (\ref{H1}). This is
clumsy and ruins the near factorization of the overlap matrix. With
downfolding, the power-series expression (\ref{H2}) for the LMTO Hamiltonian
in the L\"{o}wdin-orthonormalized basis does therefore not apply.

Most LMTO calculations include non-ASA corrections to the Hamiltonian and
overlap matrices, such as the combined correction for the neglected
integrals over the interstitial region and the neglected partial waves of
high $l.$ This brings in the first energy derivative of the structure
matrix, ${\sf \dot{S},}$ in a way which makes the formalism clumsy.\cite
{An,Kanpur} Our current, second-generation LMTO code\cite{LMTO47} is useful
and quite accurate for calculating energy bands because it includes
downfolding in addition to the combined correction,\cite{Z} but the
underlying formalism is so complicated that we never tried to publish it. On
the other hand, the combined correction is often important, and so is
downfolding because it is the only accurate means of avoiding ''ghost
bands''. The reason for the lost elegance beyond the ASA is that, whereas
the LMTO basis is complete to first order in $\varepsilon -\varepsilon _\nu $
{\em inside} the spheres, it is only complete to zeroth order in the {\em %
interstitial}. A compact formalism is therefore obtained only when the
interstitial region is neglected, and that is what the ASA does, simply by
substituting the MT spheres by space-filling spheres and neglecting the
overlap errors. The proof that the KKR equations hold to leading order for
overlapping potentials\cite{Boston} does not apply to the LMTO-ASA formalism.

There {\em are} LMTO methods sufficiently {\em accurate} to provide {\it ab
initio} structural energies and forces within density-functional theory.\cite
{FP-LMTO} For the reason mentioned above, the LMTOs for such methods\cite
{Vitos} are defined with respect to non-overlapping potentials, and since
there is considerable probability that a valence or conduction electron is
in the interstitial region, {\em outside} atom-centered, non-overlapping
spheres, an accurate basis has to include extra degrees of freedom to
describe this region, empty-sphere orbitals centered at interstitial sites
and/or atom-centered LMTOs with tails of different kinetic energies
(multiple kappa$\,$-sets). Moreover, these methods do not use small and
short-ranged representations. Finally, since a non-overlapping MT potential
is a poor approximation to the self-consistent potential, these methods {\em %
must} include the matrix elements of the full potential. Hence, the
formalisms are set up to provide final, numerical results and by themselves
provide little insight.

\medskip

\noindent \underline{Third-generation LMTOs}

\smallskip

In this paper we shall modify the LMTO set without increasing its size, in
such a way that it becomes complete to first order{\em \ }in the
interstitial region too. This is a rather natural thing to do, once the
screened spherical waves have been defined in terms of hard-sphere radii.
For the MT Hamiltonian, including downfolding, we shall regain the simple
formulas from the ASA, provided that $\left| \phi \right\rangle ,$ $h,$ $o,$
and $p$ are suitably redefined. The Hamiltonian and overlap matrices are now
given {\em solely} in terms of the screened and renormalized KKR matrix,
which we shall name $K\left( \varepsilon _\nu \right) ,$ and its first three
energy derivatives, $\dot{K}\left( \varepsilon _\nu \right) ,\;\ddot{K}%
\left( \varepsilon _\nu \right) ,\;$and $\stackrel{...}{K}\left( \varepsilon
_\nu \right) ;$ the potential parameters and the structure matrix do not
occur individually as in the formalisms of the previous generations.
Third-generation LMTOs\cite{Trieste} thus {\em do} satisfy the definition
that they form a basis constructed to reproduce the wave functions, $\Psi
_i\left( {\bf r}\right) ,$ for a MT potential to linear order in the
deviation of the single-particle energy, $\varepsilon _i,$ from a freely
chosen level, $\varepsilon _\nu .$ That is, the error of the wave functions
is of second order in $\varepsilon _i-\varepsilon _\nu $ and the error of
the single-particle energy is then of fourth order. When we use potential
wells that overlap, the wave functions will be correct to linear order in
the potential overlap and the energy error will be of second order. As we
shall demonstrate, this will remedy all shortcomings mentioned above for the
previous LMTO generations.

We shall only be concerned with solving Schr\"{o}dinger's equation in the
present paper and leave our LMTO-like expansion of the charge density,
solution of Poisson's equation, and evaluation of the total energy and
forces for future papers.\cite{Ras}

We start with a concise yet self-contained derivation of the screened KKR
method, which will lead to suitably renormalized versions of Eq.s (\ref{trf1}%
). Then we derive an expression for the error caused by using this method
for potential wells which overlap, and find that the error is of second
order in the overlap. The weak energy dependence and short range of the
screened and renormalized KKR matrix, $K\left( \varepsilon \right) ,$ is
exploited by using it to generate few-orbital, low-energy, possibly
orthonormal and short-ranged Hamiltonians for a generic high-temperature
superconductor (HTSC). Thereafter we derive the new LMTO method and
demonstrate by application to free electrons that its energy errors are
really of fourth order in $\varepsilon _i-\varepsilon _\nu .$ The power and
flexibility of the new method is demonstrated by deriving for the HTSC and
for diamond-structured silicon various LMTO sets. Using non-orthogonal $sp^3$
sets for Si, we can get an accurate first-principles description of the
valence {\em and} conduction bands if a 12th-nearest-neighbor range is
allowed in the Hamiltonian and overlap matrices. With $\varepsilon _\nu $
chosen in the middle of the valence band, a 6th-n.n. $sp^3$-set suffices for
an accurate description of the valence band and a reasonable description of
the conduction band. In order to halve the number of matrix elements, even
for a non-orthogonal basis, we use a formalism analogous to (\ref{H1}) where
the off-diagonal elements of $o$ and $p$ have been neglected so that $h$ is
the only matrix. With this simplification of the Hamiltonian and overlap
matrices, retaining the 6th-n.n. $sp^3$ basis and the low $\varepsilon _\nu ,
$ the description of the valence band remains good and merely the conduction
band deteriorates. Finally, it is possible to limit the range to 3rd nearest
neighbors provided that $d$-orbitals are included in the basis. In the last
section, we demonstrate that not only for the KKR method, but also for the
new LMTO method, the overlap error is of second order and that this can be
exploited to get completely rid of the empty-sphere wells in the diamond
structure.

\bigskip

\noindent
SCREENED SPHERICAL WAVES

\smallskip

We start by defining sets of solutions of the wave equation, $\left[ \Delta
+\varepsilon \right] \psi \left( \varepsilon ,{\bf r}\right) =0,$ so-called
screened-spherical-wave sets, $\left\{ \psi _{RL}^a\left( \varepsilon ,{\bf %
r-R}\right) \right\} ,$ which will serve as interstitial (envelope)
functions for the basis that we shall use for solving Schr\"{o}dinger's
equation. The members of a screened-spherical-wave (SSW) set are obtained by
letting $R$ run over all atomic sites and $L$ over all angular-momenta for
which the scattering is strong. The set is labelled by the superscript $a.$
Instead of defining $\psi _{RL}^a\left( \varepsilon ,{\bf r}_R\right) $ as a
specific linear combination of spherical Neumann functions like in Eq. (\ref
{trf1}), we specify it in terms of an inhomogeneous boundary condition which
is illustrated in Fig.s \ref{Boundary} and \ref{SSW} and is given as follows:

\input{infigBoundary.tek}

Concentric with each MT sphere, $R^{\prime },$ we imagine a series of
possibly coinciding {\em ''hard''} spheres with radii $a_{R^{\prime
}L^{\prime }}.$ Now, $\psi _{RL}^a\left( \varepsilon ,{\bf r}_R\right) $ is
that solution of the wave equation whose $Y_{R^{\prime }L^{\prime }}\left( 
{\bf \hat{r}}_{R^{\prime }}\right) $ projection on the $R^{\prime }L^{\prime
}$ sphere equals $\delta _{RL,R^{\prime }L^{\prime }},$ that is, 1 on its
own sphere and 0 on all other spheres. We do not associate SSWs and hard
spheres with {\em weakly-} and {\em non-}scattering channels. For such a
channel, the $Y_{R^{\prime }L^{\prime }}\left( {\bf \hat{r}}_{R^{\prime
}}\right) $ projection of the SSWs is defined to be a regular solution of
the corresponding radial Schr\"{o}dinger equation, that is, it matches onto
the irregular wave-equation solution $j_{l^{\prime }}\left( \kappa
r_{R^{\prime }}\right) -\tan \eta _{R^{\prime }l^{\prime }}\left( \kappa
\right) \,n_{l^{\prime }}\left( \kappa r_{R^{\prime }}\right) ,$ times some
constant, $c_{R^{\prime }L^{\prime },RL}^a\left( \varepsilon \right) .$ The
weakly- and non-scattering channels are thus parts of the SSW and will not
enter the screened KKR- and LMTO matrices explicitly. All high-$l^{\prime }$
channels are non-scatterers $\left[ \tan \eta _{R^{\prime }l^{\prime
}}\left( \kappa \right) =0\right] $ due to the dominance of the centrifugal
barrier. Empty spheres are examples of a weak scatterers. {\em Strong}
scatterers are then, by definition, those channels with which we associate
SSWs and hard spheres. Note that all SSWs in the set have the same boundary
condition, except for the $\delta _{RL,R^{\prime }L^{\prime }}.$ The SSW
set, $\left\{ \psi _{RL}^a\left( \varepsilon ,{\bf r}_R\right) \right\}
\equiv \left| RL\right\rangle \left\langle {\bf r}\right| ,$ may thus be
considered as an unperturbed Green function in a hybrid representation.

Fig. \ref{SSW} shows an SSW for the hypothetical case of only strong
scattering. Weak- and non-scattering channels would have shown up as little
tails extending into the two hard spheres. Such tails may be seen in Fig. 
\ref{KPW} where the dashed curve is an SSW for Si. In this figure we have
set the radial functions of the strongly scattering channels to zero inside
the hard spheres and, defined in this way, $\psi _{RL}^a\left( \varepsilon ,%
{\bf r}_R\right) $ jumps by the amount $Y_L\left( {\bf \hat{r}}_R\right) $
at its own hard sphere, $r_R{\rm =}a_{RL},$ and has kinks at all hard
spheres. Had we instead chosen to continue also the strongly scattering
channels of the SSW into the hard spheres, the SSW would have been smooth,
but diverging at the sites of the strongly scattering atoms, each radial
part going as $j_{l^{\prime }}\left( \kappa r_{R^{\prime }}\right) -\tan
\alpha _{R^{\prime }L^{\prime }}\left( \kappa \right) n_{l^{\prime }}\left(
\kappa r_{R^{\prime }}\right) .$ In order to get more feeling for SSWs, let
us consider some limiting cases:

If we specify $a_{RL}=a\rightarrow 0$ for all channels, we obtain the {\em %
bare} spherical waves. These are known analytically but, except maybe for
small molecules, we never use them. Nevertheless, with the normalization
specified above they are:\cite{Trieste} 
\begin{eqnarray*}
\psi _{0L}^0\left( \varepsilon ,{\bf r}\right) &=&-\frac{\left( \kappa
a\right) ^{l+1}}{\left( 2l-1\right) !!}n_l\left( \kappa r\right) Y_L\left( 
{\bf \hat{r}}\right) =\left[ \frac ar\right] ^{l+1}\left[ 1+\frac{%
\varepsilon r^2}{2\left( 2l-1\right) }-...\right] Y_L\left( {\bf \hat{r}}%
\right) \\
&\rightarrow &\frac{\left( \kappa a\right) ^{l+1}}{\left( 2l-1\right) !!}%
\frac{\cos \left( \kappa r-l\pi /2\right) }{\kappa r}Y_L\left( {\bf \hat{r}}%
\right) ,\quad {\rm for\quad }r\rightarrow \infty ,
\end{eqnarray*}
when $\varepsilon \ge 0,$ and where $\left( 2l-1\right) !!\equiv \left(
2l-1\right) \left( 2l-3\right) ..\cdot 1$ and $\left( -1\right) !!\equiv -1.$
When $\varepsilon \le 0:$%
\begin{eqnarray*}
\psi _{0L}^0\left( \varepsilon ,{\bf r}\right) &=&-\frac{\left( \kappa
a\right) ^{l+1}}{\left( 2l-1\right) !!}\left[ n_l\left( \kappa r\right)
+ij_l\left( \kappa r\right) \right] Y_L\left( {\bf \hat{r}}\right) \\
&=&\left[ \frac ar\right] ^{l+1}\left\{ \left[ 1+\frac{\varepsilon r^2}{%
2\left( 2l-1\right) }-\right] -\frac{\left( \varepsilon r^2\right) ^lr\sqrt{%
-\varepsilon }}{\left( 2l+1\right) \left[ \left( 2l-1\right) !!\right] ^2}%
\left[ 1-\frac{\varepsilon r^2}{2\left( 2l+3\right) }+\right] \right\}
Y_L\left( {\bf \hat{r}}\right) \\
&\rightarrow &\frac{a^{l+1}\left( \sqrt{-\varepsilon }\right) ^l}{\left(
2l-1\right) !!}\frac{\exp \left( -r\sqrt{-\varepsilon }\right) }rY_L\left( 
{\bf \hat{r}}\right) ,\quad {\rm for\quad }r\rightarrow \infty .
\end{eqnarray*}
These expressions hold only for $r\gg a.$ Unlike screened spherical waves,
the bare ones are eigenfunctions of angular momentum and are independent of
the surroundings. For positive energies they have long range. Like all
screened spherical waves, the normalization of the bare ones is such that
they are dimensionless and, unlike the Bessel and Neumann functions, they
depend little on energy near the hard spheres.

Another case is when we specify $a_{RL}{\rm =}a_R$ for {\em all }$L,$ and
take $\varepsilon {\rm =}0.$ Then $\psi _{0L}^a\left( 0,{\bf r}\right) $ is
proportional to the electrostatic potential from a $2^l$-pole at the origin,
surrounded by grounded conducting spheres with radii $a_R$ centered at the
other sites. Since the hard spheres at the neighbors break the spherical
symmetry around the origin, the SSW has pure angular-momentum character
merely at its own sphere, and this holds only as long as the own sphere
coincides with all other spheres concentric with it. Changing the energy
will not change the SSW much. If we now let $a_{R^{\prime }L^{\prime }}$ be
zero for high $l^{\prime }$'s, the SSWs will ''wobble'' into the hard
spheres.

If the hard-sphere radii are generated from repulsive potential wells,\cite
{Zeller} the SSWs are the ''impurity states'' for that repulsive MT
potential.

Since the strongly-scattering components of the SSWs are forced to vanish at
all{\em \ }surrounding hard spheres, the {\em range} of the SSWs depends on
the choice of hard spheres and energy: Consider the spectrum $\varepsilon
_i^a$ of the wave equation with the {\em homogeneous} boundary condition
that the solutions vanish at {\em all} spheres. This spectrum has a
continuum starting at $\varepsilon _c^a,$ which in the absence of screening
is at zero and which rises with increasing hard-sphere radii. Now, the SSWs
are {\em localized }or {\em delocalized} depending on whether their energy
is below or above the bottom of the continuum. Since we choose
energy-independent boundary conditions for the SSWs, their {\em energy
dependence} merely enters through the wave equation, that is through their
curvature, and is therefore small when the wavelength exceeds the diameter
of the largest interstitial in the hard-sphere solid.

If all hard spheres centered on the same site would coincide, then the hard
spheres would have to be {\em smaller than touching} because, if two spheres
had a point (or a circle) in common, then each one of the SSWs centered on
the two spheres would be required to be both zero and non-zero at that point
(or circle). When only a few low-$l$ channels scatter strongly, neighboring
hard spheres may intersect. With decreasing hard-sphere interstitial, the
SSW sets thus in general become more and more localized, until the hard
spheres start to intersect. Since from there on, the SSWs are forced to
change rapidly near the common circles, their behavior becomes chaotic as
the circles grow.

We shall generate the screened spherical waves from the bare ones, because
those are the only ones we know analytically. Hence, we first consider the
question of how to expand an arbitrary wave-equation solution, $\Psi \left(
\varepsilon ,{\bf r}\right) ,$ which is regular in all space, except
possibly at the atomic sites, in an SSW set, $\left\{ \psi _{RL}^a\left(
\varepsilon ,{\bf r}_R\right) \right\} ,$ with the same energy. If the
number of atoms is finite, this energy is supposed to be negative. Moreover,
since $\Psi \left( \varepsilon ,{\bf r}\right) $ is a solution of the wave
equation, the SSW set is supposed to have no weakly-scattering channels and, 
{\it a priory, }we treat all channels as strong scatterers. Finally, we
shall not truncate the SSWs inside the hard spheres, but let them continue
to the centers. We now expand $\Psi \left( \varepsilon ,{\bf r}\right) $ in
spherical harmonics on the hard spheres of the SSW set, thus obtaining the
coefficients, $\Psi _{Rlm}\left( \varepsilon ,a_{Rlm}\right) .$ Unless all
of these vanish, the linear combination converges to $\Psi \left(
\varepsilon ,{\bf r}\right) :$ 
\[
\lim_{\lambda _R\rightarrow \infty }\sum_R\sum_{l=0}^{\lambda
_R}\sum_{m=-l}^l\psi _{Rlm}^a\left( \varepsilon ,{\bf r}_R\right) \,\Psi
_{Rlm}\left( \varepsilon ,a_{Rlm}\right) =\Psi \left( \varepsilon ,{\bf r}%
\right) 
\]
because, by construction, the linear combination is a solution of the wave
equation with the proper energy, and this solution matches $\Psi \left(
\varepsilon ,{\bf r}\right) $ channel by channel. In order to convince
oneself that the latter is sufficient, one may start repeating the argument
using an SSW set with $L$-independent hard spheres. In that case, $\Psi
\left( \varepsilon ,{\bf r}\right) $ coincides with the linear combination
on a closed boundary, because in the case where the system is infinite such
a boundary is formed by the entity of all hard spheres, and in the case
where the system is finite, the boundary is formed by the hard spheres plus
the infinity, where both $\Psi \left( \varepsilon ,{\bf r}\right) $ and the
linear combination vanish since the energy is negative.

If all the coefficients $\Psi _{Rlm}\left( \varepsilon ,a_{Rlm}\right) $
vanish then $\Psi \left( \varepsilon ,{\bf r}\right) $ is an eigenfunction
of the hard-sphere solid. In this case a complete set must include, in
addition to the SSWs, the degenerate eigenfunctions, or we may choose a
different SSW set for the expansion of $\Psi \left( \varepsilon ,{\bf r}%
\right) .$

Changing the hard-sphere radii, but not the sites and the energy, produces
another set of SSW's which is also complete in the above-mentioned sense.
All such sets are therefore linearly dependent. A set of hard-sphere radii
is said to specify a {\em representation} and the transformation from the $a$
to the $b$ representation is obtained by substituting $\psi _{R^{\prime
}L^{\prime }}^b\left( \varepsilon ,{\bf r}_{R^{\prime }}\right) $ for $\Psi
\left( \varepsilon ,{\bf r}\right) $ in the above. Hence, the transformation
is 
\begin{equation}
\psi _{R^{\prime }L^{\prime }}^b\left( \varepsilon ,{\bf r}_{R^{\prime
}}\right) =\sum_{RL}\psi _{RL}^a\left( \varepsilon ,{\bf r}_R\right) \,\psi
_{RL,R^{\prime }L^{\prime }}^b\left( \varepsilon ,a_{RL}\right) ,
\label{SSWtrf}
\end{equation}
where $\psi _{RL,R^{\prime }L^{\prime }}^b\left( \varepsilon ,a_{RL}\right) $
are the $RL$ components at $a_{RL}$-spheres of the functions $\psi
_{R^{\prime }L^{\prime }}^b\left( \varepsilon ,{\bf r}_{R^{\prime }}\right)
. $ Expanding now the left and right-hand sides in spherical harmonics on
the $b_{R^{\prime \prime }L^{\prime \prime }}$-spheres we obtain: 
\begin{equation}
\delta _{R^{\prime \prime }R^{\prime }}\delta _{L^{\prime \prime }L^{\prime
}}=\sum_{RL}\psi _{R^{\prime \prime }L^{\prime \prime },RL}^a\left(
\varepsilon ,b_{R^{\prime \prime }L^{\prime \prime }}\right) \,\psi
_{RL,R^{\prime }L^{\prime }}^b\left( \varepsilon ,a_{RL}\right) .
\label{complete}
\end{equation}
The two matrices $\psi ^a\left( \varepsilon ,b\right) $ and $\psi ^b\left(
\varepsilon ,a\right) $ are thus each others inverses.

\bigskip

\noindent SLOPE AND STRUCTURE MATRICES

\smallskip

We have specified the SSWs by their nodes and shall need their radial
derivatives at the hard spheres, that is, the dimensionless {\em slope
matrix.} Its element $S_{R^{\prime }L^{\prime },RL}^a\left( \varepsilon
\right) $ is defined as $a_{R^{\prime }L^{\prime }}$ times the $L^{\prime }$%
-component of the radial derivative at the $a_{R^{\prime }L^{\prime }}$%
-sphere, with the positive direction taken outwards from ${\bf R}^{\prime },$
of $\psi _{RL}^a\left( \varepsilon ,{\bf r}_R\right) .$ This is illustrated
in Fig. \ref{SSW}.

In fact knowledge of the hard spheres and the slope matrix makes generation
of the SSW set a simple matter: The{\em \ }spherical-harmonics expansion
around any site, ${\bf R}^{\prime },$ of any member, $\psi _{RL}^a\left(
\varepsilon ,{\bf r}_R\right) $, of the set is given by radial functions and
the function for the $L^{\prime }$ channel is: 
\begin{equation}
\ \psi _{R^{\prime }L^{\prime },RL}^a\left( \varepsilon ,r_{R^{\prime
}}\right) =\ f_{l^{\prime }}\left( \varepsilon ,a_{R^{\prime }L^{\prime
}},r_{R^{\prime }}\right) \,\delta _{R^{\prime }L^{\prime },RL}+g_{l^{\prime
}}\left( \varepsilon ,a_{R^{\prime }L^{\prime }},r_{R^{\prime }}\right)
\,S_{R^{\prime }L^{\prime },RL}^a\left( \varepsilon \right) .  \label{local}
\end{equation}
The local expansion converges for $r_{R^{\prime }}$ smaller than the
distance to the nearest site. In (\ref{local}), $f_l$ and $g_l$ are
solutions of the radial wave equation, $\left[ d^2/dr^2-l\left( l+1\right)
/r^2+\varepsilon \right] rf_l=0,$ with the following boundary conditions for 
$r{\rm =}a:\;$ $f$ has value one\ and slope zero,{\em \ }and $g$ has value
zero\ and slope $1/a.$

The SSW-set, $\left| \psi ^a\left( \varepsilon \right) \right\rangle ,$ may
also be expressed {\em globally} as a linear combination of some known set, $%
\left| \psi ^b\left( \varepsilon \right) \right\rangle ,$ as we saw in Eq. (%
\ref{SSWtrf}). The transformation matrix, $\psi ^a\left( \varepsilon
,b\right) ,$ is then given by Eq. (\ref{local}) with $r_{R^{\prime }}$
substituted by $b_{R^{\prime }L^{\prime }}.$ With the use of Eq. (\ref{local}%
), the completeness relation (\ref{complete}) thus expresses the
transformation from $S^b\left( \varepsilon \right) $ to $S^a\left(
\varepsilon \right) .$

In order to {\em generate} the slope matrix, we transform to the bare set,
which is known analytically: Using Eq. (\ref{local}), $S^a\left( \varepsilon
\right) $ is expressed in terms of $\psi ^a\left( \varepsilon ,0\right) ,$
which is computed as the inverse of $\psi ^0\left( \varepsilon ,a\right) .$
The latter follows from the local, spherical-harmonics expansion about ${\bf %
R}^{\prime }$ of the Neuman function centered at ${\bf R}$ $\left( \neq {\bf %
R}^{\prime }\right) $: $\kappa n_l\left( \kappa {\bf r}_R\right) Y_L\left( 
{\bf \hat{r}}_R\right) =\sum_{L^{\prime }}j_{l^{\prime }}\left( \kappa {\bf r%
}_{R^{\prime }}\right) Y_{L^{\prime }}\left( {\bf \hat{r}}_{R^{\prime
}}\right) B_{R^{\prime }L^{\prime },RL}\left( \kappa \right) ,$ where 
\begin{equation}
B_{R^{\prime }L^{\prime },RL}\left( \kappa \right) \equiv \sum_{l"}4\pi
\,i^{-l+l^{\prime }-l^{\prime \prime }}C_{LL^{\prime }l^{\prime \prime
}}\,\kappa n_{l^{\prime \prime }}\left( \kappa \left| {\bf R-R}^{\prime
}\right| \right) \,Y_{l^{\prime \prime },m^{\prime }-m}^{*}\left( \widehat{%
{\bf R-R}^{\prime }}\right)   \label{B}
\end{equation}
is the KKR structure matrix, which is Hermitian. The summation runs over $%
l^{\prime \prime }=\left| l^{\prime }-l\right| ,\;\left| l^{\prime
}-l\right| +2,\,...,\;l^{\prime }+l,$ and $i^{-l+l^{\prime }-l^{\prime
\prime }}$ is real because $C_{LL^{\prime }L^{\prime \prime }}\equiv \int
Y_L(\hat{r})Y_{L^{\prime }}^{*}(\hat{r})Y_{L^{\prime \prime }}(\hat{r})d\hat{%
r}$. The on-site elements of $B\left( \kappa \right) $ vanish. In this way,
we obtain the most important result: 
\begin{equation}
aS^a\left( \varepsilon \right) -aD\left\{ j\left( \kappa a\right) \right\}
=\frac 1{j\left( \kappa a\right) }\left[ B\left( \kappa \right) +\kappa \cot
\alpha \left( \kappa \right) \right] ^{-1}\frac 1{j\left( \kappa a\right) },
\label{scr}
\end{equation}
where $a,\;j\left( \kappa a\right) ,$\ $D\left\{ j\left( \kappa a\right)
\right\} ,$\ and $\cot \alpha \left( \kappa \right) $ are {\em diagonal}
matrices with elements $a_{RL},$ $j_l\left( \kappa a_{RL}\right) ,$ $%
D\left\{ j_l\left( \kappa a_{RL}\right) \right\} {\rm =}\kappa
a_{RL}j^{\prime }\left( \kappa a\right) /j\left( \kappa a\right) ,$ and $%
\cot \alpha _{RL}\left( \kappa \right) .$ The quantity $a_{R^{\prime
}L^{\prime }}S_{R^{\prime }L^{\prime },RL}^a\left( \varepsilon \right) ,$
which is $a_{R^{\prime }L^{\prime }}^2$ times the $L^{\prime }$-component of
the radial derivative of $\psi _{RL}^a\left( \varepsilon \right) $ at the $%
a_{R^{\prime }L^{\prime }}$-sphere, form the elements of a matrix which is 
{\em Hermitian. }This matrix, we call the structure matrix.

For the channels to be treated as strongly scattering with the set $\left\{
\psi ^a\right\} ,$ we take $\alpha _{RL}\left( \kappa \right) $ to be the
hard-sphere phase shifts (\ref{alpha}), and for those to be treated as
weakly scattering and, thus to be {\em downfolded} into the SSWs, we take $%
\alpha _{RL}\left( \kappa \right) $ to be the real phase shifts, $\eta
_{Rl}\left( \kappa \right) .$ The non-scattering channels do not enter the
screening calculation (\ref{scr}), since they neither scatter the bare, nor
the screened set. The strongly- and weakly-scattering channels thus
contribute to the size of the matrix to be inverted and the 
strongly-scattering 
channels are the only ones which will eventually enter the 
equations for solving Schr\"{o}dinger's equation.

Instead of expressing (\ref{B}) and (\ref{scr}) in terms of the usual
spherical Bessel and Neumann functions, one could of course have divided the
factors $\kappa ^l$ and $\kappa ^{-l-1}$ out on the right-hand side of (\ref
{scr}), or used $\psi _l^0\left( \varepsilon ,r\right) $ instead of $%
n_l\left( \kappa r\right) ,$ etc.. The only difference between the last
equation of (\ref{trf1}) and equation (\ref{scr}), is that the Hermitian
matrix $aS^\alpha \left( \varepsilon \right) $ is normalized in such a way
as to make its energy dependence as small as possible, and in such a way as
to give $S^\alpha \left( \varepsilon \right) $ a geometrical interpretation,
namely as the dimensionless slope matrix. Specifically, 
\[
\kappa ^{-1}\tan \alpha \left[ B^\alpha \left( \kappa \right) -\kappa \cot
\alpha \left( \kappa \right) \right] \kappa ^{-1}\tan \alpha =\,-j\left(
\kappa a\right) a\left[ S^a\left( \varepsilon \right) -D\left\{ j\left(
\kappa a\right) \right\} \right] j\left( \kappa a\right) , 
\]
so that the screened structure matrices $B^\alpha \left( \kappa \right) $
and $aS^a\left( \varepsilon \right) $ differ because functions of energy
have been subtracted from the diagonal elements, and because the rows and
columns have been rescaled with such functions.
If we form:$\,$%
\begin{equation}
{\sf S}_{R^{\prime }L^{\prime },RL}^{{\sf a}}\left( \varepsilon \right)
\equiv -2\left( w/a_{R^{\prime }L^{\prime }}\right) ^{l^{\prime }}\left[
S_{R^{\prime }L^{\prime },RL}^a\left( \varepsilon \right) +\left( l+1\right)
\delta _{R^{\prime }L^{\prime },RL}\right] \left( w/a_{RL}\right) ^{l+1},
\label{conv}
\end{equation}
then ${\sf S}^{{\sf a}}\left( 0\right) $ is the conventional ($\kappa $=0)
LMTO structure matrix for the screening constants 
\[
{\sf a}_{RL}\left( 0\right) =\left[ 2\left( 2l+1\right) \right] ^{-1}\left(
a_{RL}/w\right) ^{2l+1}, 
\]
and ${\sf \dot{S}}^{{\sf a}}\left( 0\right) $ is its first energy derivative
for some ${\sf \dot{a}}\left( 0\right) .$ For LMTO users who have developed
a feeling for the sizes of the conventional structure constants and do not
care about the new interpretation in terms of logarithmic derivatives, it is
of course possible to use the new method in the conventional ''gauge'' (\ref
{conv}). In that case, one {\em must} substitute the old potential
functions, ${\sf P}_{RL}^{{\sf a}}\left( \varepsilon \right) ,$ by $-2\left(
w/a_{RL}\right) ^{2l+1}\left[ D\left\{ \varphi _{Rl}\left( \varepsilon
,a_{RL}\right) \right\} +l+1\right] ,$ with $D\left\{ \varphi _{Rl}\left(
\varepsilon ,a_{RL}\right) \right\} $ evaluated as explained in the
following section.\cite{Trieste}

\input{infigS.tek}

Whereas the slope matrix specifies the normal gradients on the hard spheres
of all functions in the SSW set, its first energy derivative, $\dot{S}%
_{RL,R^{\prime }L^{\prime }}^a\left( \varepsilon \right) ,$ specifies the
normal gradients of the first-energy derivative functions, $\dot{\psi}%
_{RL}^a\left( \varepsilon \right) ,$ as illustrated at the bottom of Fig. 
\ref{SSW}. Since the hard spheres are independent of energy, the
energy-derivative functions will vanish at {\em all} hard spheres, including
their own. The first energy derivative of the structure matrix in addition
gives the {\em overlap matrix} of the SSW set: $\left\langle \psi
_{RL}^a\left( \varepsilon \right) |\psi _{R^{\prime }L^{\prime }}^a\left(
\varepsilon \right) \right\rangle =a\dot{S}_{RL,R^{\prime }L^{\prime
}}^a\left( \varepsilon \right) .$ This equation follows from the more
general one: $\left\langle \psi _{RL}^a\left( \varepsilon \right) |\psi
_{R^{\prime }L^{\prime }}^a\left( \varepsilon ^{\prime }\right)
\right\rangle =\,a_{RL}\left[ S_{RL,R^{\prime }L^{\prime }}^a\left(
\varepsilon \right) -S_{RL,R^{\prime }L^{\prime }}^a\left( \varepsilon
^{\prime }\right) \right] /\left( \varepsilon -\varepsilon ^{\prime }\right)
,$ which may be derived by use of Green's second theorem.\cite{Trieste}
Here, the strongly scattering radial components have been truncated inside
the corresponding hard spheres as illustrated in Fig.s \ref{Boundary} and 
\ref{SSW}, while the remaining, regular components extend to the centers of
the spheres.

Considered as functions of $\varepsilon ,$ the eigenvalues of the structure
matrix $aS^a\left( \varepsilon \right) $ have poles when $\varepsilon $
coincides with an energy eigenvalue, $\varepsilon _i^a,$ of the hard-sphere
solid. For practical purposes, the $a$-radii can be chosen in such a way
that the energies of interest to us are well below the bottom of the
hard-sphere continuum, and below any localized state of the hard-sphere
solid. For such energies, the eigenvalues of $aS^a\left( \varepsilon \right) 
$ are analytical functions of $\varepsilon $. This latter point is
demonstrated in the left-hand side of Fig. \ref{S}, which also demonstrates
that the energy dependence is weak over the $\pm 10$ eV region considered.
The right-hand side shows that, for low energies or close sphere packings,
the slope matrix decays by an order of magnitude per shell of neighbors. For
a monotonically decaying SSW we expect, as illustrated in Fig. \ref{SSW}, a
negative slope at its own hard sphere and positive slopes at the neighboring
spheres. This is also the behavior found in Fig. \ref{S}, at least
throughout the first three shells.

In conclusion, the slope matrix generated by inversion of the non singular
matrix (\ref{scr}) contains all the information we shall need about the SSW
set.

\bigskip

\noindent SOLVING SCHR\H{O}DINGER'S EQUATION WITH KINKED PARTIAL WAVES

\smallskip

We now come to consider Schr\"{o}dinger's equation, $\left[ -\Delta +V\left( 
{\bf r}\right) -\varepsilon _i\right] \Psi _i\left( {\bf r}\right) =0,$ for
a MT potential and begin by showing that with our screened spherical waves
it is a rather simple matter to formulate the matching problem for the
solutions, $\Psi _i\left( {\bf r}\right) ,$ algebraically:

First we integrate the radial Schr\"{o}dinger equation for each {\em strongly%
} scattering channel {\em outwards} from the origin to the MT radius $s_R$
in the potential well $v_R\left( r_R\right) ,$ and then {\em inwards} in 
{\em zero} potential (the MT zero) from $s_R$ to the hard-sphere radius $%
a_{RL}$. The outwards integration yields the radial partial wave $\phi
_{Rl}\left( \varepsilon ,r_R\right) ,$ and the subsequent inwards
integration yields the radial partial wave ''as seen from free space'' $%
\varphi _{Rl}\left( \varepsilon ,r_R\right) ,$ with radial logarithmic
derivative $D\left\{ \varphi _{Rl}\left( \varepsilon ,a_{RL}\right) \right\} 
$ at the hard sphere. These two waves match continuously and differentiably
at $s_R$ and they may be seen in the left-hand side of Fig. \ref{KPW}, after
multiplication by $Y_L\left( {\bf \hat{r}}_R\right) .$ Let us assume that $%
\phi _{RL}^a\left( \varepsilon ,r_R\right) $ and $\varphi _{RL}^a\left(
\varepsilon ,r_R\right) $ have been normalized in such a way that $\varphi
_{RL}^a\left( \varepsilon ,a_{RL}\right) \equiv 1$ at the hard sphere; this
is what the superscript $a$ here indicates. In the case where the hard
spheres have been chosen to depend on $m,$ radial functions of the same $Rl$
may have different normalizations, hence the subscript $L$ rather than $l.$
With this normalization, the free partial wave matches continuously, but
with the kink $S_{RL,RL}^a\left( \varepsilon \right) -D\left\{ \varphi
_{Rl}\left( \varepsilon ,a_{RL}\right) \right\} $, to the $RL$-projection,$\
\psi _{R^{\prime }L^{\prime },RL}^a\left( \varepsilon ,r_{R^{\prime
}}\right) ,$ of the corresponding SSW, $\psi _{RL}^a\left( \varepsilon ,{\bf %
r}_R\right) .$ Let us furthermore truncate $\phi _{RL}^a\left( \varepsilon
,r_R\right) $ and $\varphi _{RL}^a\left( \varepsilon ,r_R\right) $ outside
the MT sphere $\left( 0|s_R\right) $ and, like the SSW, let us truncate $%
\varphi _{RL}^a\left( \varepsilon ,r_R\right) $ also inside the $a_{RL}$%
-sphere. The function $\left[ \phi _{RL}^a\left( \varepsilon ,r_R\right)
-\varphi _{RL}^a\left( \varepsilon ,r_R\right) \right] Y_L\left( {\bf \hat{r}%
}_R\right) $ thus equals the proper partial wave inside the hard sphere,
where it jumps by $-Y_L\left( {\bf \hat{r}}_R\right) ,$ and it vanishes
quadratically at the MT sphere with a prefactor proportional to the MT
discontinuity $v_R\left( s_R\right) .$ To this function we now add the
corresponding SSW thus obtaining the {\em kinked partial wave} (KPW): 
\begin{equation}
\Phi _{RL}^a\left( \varepsilon ,{\bf r}_R\right) \equiv \left[ \phi
_{RL}^a\left( \varepsilon ,r_R\right) -\varphi _{RL}^a\left( \varepsilon
,r_R\right) \right] Y_L\left( {\bf \hat{r}}_R\right) +\psi _{RL}^a\left(
\varepsilon ,{\bf r}_R\right) ,  \label{KinkPW}
\end{equation}
which is also shown in Fig. \ref{KPW}. This function is everywhere
continuous, but has kinks of size $S_{R^{\prime }L^{\prime },RL}^a\left(
\varepsilon \right) -D\left\{ \varphi _{Rl}\left( \varepsilon ,a_{RL}\right)
\right\} \delta _{R^{\prime }L^{\prime },RL}$ at the hard $a_{R^{\prime
}L^{\prime }}$-spheres.

At such a sphere, the kink of the {\em linear combination\ }of KPWs, $%
\sum_{RL}\Phi _{RL}^a\left( \varepsilon ,{\bf r}_R\right) c_{RL}^a\left(
\varepsilon \right) ,$ is therefore $\sum_{RL}\left[ S_{R^{\prime }L^{\prime
},RL}^a\left( \varepsilon \right) -D\left\{ \varphi _{Rl}\left( \varepsilon
,a_{RL}\right) \right\} \delta _{R^{\prime }L^{\prime },RL}\right]
\,c_{RL}^a\left( \varepsilon \right) .$ If we can now find an energy, $%
\varepsilon _i,$ and coefficients, $c_{RL,i}^a,$ such that 
\begin{equation}
\sum_{RL}\left[ S_{R^{\prime }L^{\prime },RL}^a\left( \varepsilon _i\right)
-D\left\{ \varphi _{Rl}\left( \varepsilon _i,a_{RL}\right) \right\} \delta
_{R^{\prime }L^{\prime },RL}\right] \,c_{RL,i}^a=0\quad \quad {\rm for\;all\;%
}R^{\prime }L^{\prime },  \label{KKR}
\end{equation}
then the corresponding linear combination is {\em smooth} and therefore
solves Schr\"{o}dinger's equation with $\varepsilon _i$ as an energy
eigenvalue.

\input{infigKPW.tek}

The statement that the {\em ''kink-cancellation condition''} (\ref{KKR})
leads to a solution of Schr\"{o}dinger's equation is exact only for a
non-overlapping MT potential. Before continuing to the case of overlapping
potentials, let us scrutinize our proof a little closer. Each KPW is
constructed to be a solution of Schr\"{o}dinger's equation at energy $%
\varepsilon ,$ except in all shells between concentric MT- and hard spheres,
and except for the kinks at the hard spheres. In the case where we choose
all concentric hard spheres to coincide with the MT sphere $\left( a_{RL}%
{\rm =}s_R\right) ,$ all shells vanish so the statement is obviously true.
That it holds also when the hard spheres are different from the concentric
MT sphere, follows from the fact that for a linear combination with all
kinks cancelled, each $\varphi _{R^{\prime }L^{\prime }}^a\left( \varepsilon
_i,r_{R^{\prime }}\right) c_{R^{\prime }L^{\prime },i}^a$ matches the $%
R^{\prime }L^{\prime }$-projection, $\sum_{RL}\psi _{R^{\prime }L^{\prime
},RL}^a\left( \varepsilon _i,r_{R^{\prime }}\right) c_{RL,i}^a,$ of the
linear combination of SSWs, $\sum_{RL}\psi _{RL}^a\left( \varepsilon _i,{\bf %
r}_R\right) c_{RL,i}^a,$ at $a_{R^{\prime }L^{\prime }}$ in value{\em \ and
in slope,} and since both radial functions are solutions of the {\em same}
second-order differential equation, namely the $l^{\prime }$'th radial wave
equation, they must be {\em identical.} As a consequence, 
\begin{equation}
\varphi _{R^{\prime }L^{\prime }}^a\left( \varepsilon _i,r_{R^{\prime
}}\right) c_{R^{\prime }L^{\prime },i}^a-\sum_{RL}\psi _{R^{\prime
}L^{\prime },RL}^a\left( \varepsilon _i,r_{R^{\prime }}\right)
c_{RL,i}^a=0,\;{\rm for\;}0\le r_{R^{\prime }}\le s_{R^{\prime }}\;{\rm and\;%
}R^{\prime }L^{\prime }\in {\rm strong\;scat.}  \label{canc}
\end{equation}
Inside the $s_{R^{\prime }}$-sphere then, only terms which satisfy Schr\"{o}%
dinger's equation remain, namely $Y_{L^{\prime }}\left( {\bf \hat{r}}%
_{R^{\prime }}\right) \sum_{RL}\phi _{R^{\prime }L^{\prime },RL}^a\left(
\varepsilon _i,r_{R^{\prime }}\right) c_{RL,i}^a$ and the weakly- and
non-scattering channels of $\sum_{RL}\psi _{RL}^a\left( \varepsilon _i,{\bf r%
}_R\right) c_{RL,i}^a.$

\input{infigOVL.tek}

The KPW is defined in (\ref{KinkPW}) as the SSW plus the central, pure
angular-momentum contribution $\phi -\varphi ,$ which vanishes quadratically
at the MT sphere. In analogy with Slater's augmented plane wave (APW), the
KPW might have been named an augmented screened spherical wave. This analogy
is only complete though when all hard spheres coincide with their concentric
MT sphere.

Next we consider the case of MT {\em overlap}. Suppose that we have solved
the kink-cancellation equations (\ref{KKR}) with logarithmic derivatives
calculated for potential wells which overlap. To what extent is the
resulting {\em smooth }function, $\Psi _i\left( {\bf r}\right) \equiv
\sum_{RL}\Phi _{RL}\left( \varepsilon _i,{\bf r}_R\right) c_{RL,i},$ a
solution of Schr\"{o}dinger's equation for the superposition of these
overlapping wells? The situation is sketched in Fig. \ref{OVL} and the
answer is, that the smooth superposition of KPWs solves Schr\"{o}dinger's
equation to leading (first) order in the potential overlap.

Since we have only considered the strongly-scattering channels in this
one-dimensional figure, let us now be a bit more careful. Using the
definition (\ref{KinkPW}) and the following definitions: $\psi _i\left( {\bf %
r}\right) \equiv \sum_{RL}\psi _{RL}^a\left( \varepsilon _i,{\bf r}_R\right)
c_{RL,i},\;\phi _R^a\left( \varepsilon _i,{\bf r}_R\right) \equiv \sum_L\phi
_{RL}^a\left( \varepsilon _i,{\bf r}_R\right) c_{RL,i},$ and similarly for $%
\varphi _R^a\left( \varepsilon _i,{\bf r}_R\right) ,$ we obtain: 
\[
\left[ -\Delta +\sum_Rv_R\left( r_R\right) -\varepsilon _i\right] \Psi
_i\left( {\bf r}\right) =\sum_{R^{\prime }}v_{R^{\prime }}\left(
r_{R^{\prime }}\right) \sum_{R\neq R^{\prime }}\left[ \phi _R^a\left(
\varepsilon _i,{\bf r}_R\right) -\varphi _R^a\left( \varepsilon _i,{\bf r}%
_R\right) \right] 
\]
\[
+\sum_R\left[ -\Delta +v_R\left( r_R\right) -\varepsilon _i\right] \left[
\phi _R^a\left( \varepsilon _i,{\bf r}_R\right) -\varphi _R^a\left(
\varepsilon _i,{\bf r}_R\right) \right] +\left[ -\Delta +\sum_Rv_R\left(
r_R\right) -\varepsilon _i\right] \psi _i\left( {\bf r}\right) = 
\]
\[
\sum_{R^{\prime }}v_{R^{\prime }}\left( r_{R^{\prime }}\right) \sum_{R\neq
R^{\prime }}\left[ \phi _R^a\left( \varepsilon _i,{\bf r}_R\right) -\varphi
_R^a\left( \varepsilon _i,{\bf r}_R\right) \right] -\sum_Rv_R\left(
r_R\right) \left[ \varphi _R^a\left( \varepsilon _i,{\bf r}_R\right) -\psi
_i\left( {\bf r}\right) \right] -\left[ \Delta +\varepsilon _i\right] \psi
_i\left( {\bf r}\right) 
\]
\begin{eqnarray}
&=&\sum_{R^{\prime }}v_{R^{\prime }}\left( r_{R^{\prime }}\right)
\sum_{R\neq R^{\prime }}\left[ \phi _R^a\left( \varepsilon _i,{\bf r}%
_R\right) -\varphi _R^a\left( \varepsilon _i,{\bf r}_R\right) \right]
\label{ovl1} \\
&=&\frac 12\sum_{R^{\prime }}v_{R^{\prime }}\left( r_{R^{\prime }}\right)
\sum_{R\neq R^{\prime }}v_R\left( s_R\right) \left[ \left( s_R-r_R\right)
^2+o\left( \left( s_R-r_R\right) ^2\right) \right] \phi _R^a\left(
\varepsilon _i,{\bf r}_R\right)  \label{ovl2} \\
&\sim &\frac 12\sum_{RR^{\prime }}^{{\rm pairs}}v_{R^{\prime }}\left(
r_{R^{\prime }}\right) \left[ \left( s_{R^{\prime }}-r_{R^{\prime }}\right)
^2+\left( s_R-r_R\right) ^2\right] v_R\left( s_R\right) \,\Psi _i\left( {\bf %
r}\right)  \label{ovl3}
\end{eqnarray}
Here, we have first of all made use of the fact that $\Psi _i\left( {\bf r}%
\right) $ is smooth so that we can apply the $\Delta $-operator to its
individual, kinked or discontinuous parts without keeping track of all the
resulting diverging terms, because they will cancel in the end. In obtaining
the 3rd line, we have used that $\phi $ solves Schr\"{o}dinger's equation
for its own well. Eq. (\ref{ovl1}) has then been obtained by use of Eq. (\ref
{canc}) for the strongly-scattering partial waves, plus the fact that the
weakly- and non-scattering channels $\left( \Lambda \right) $ of $\psi
_i\left( {\bf r}\right) $ solve Schr\"{o}dinger's equation, {\it i.e.} that $%
\left[ \Delta +\varepsilon \right] \psi _{RL}^a\left( \varepsilon ,{\bf r}%
_R\right) =\,\sum_{R^{\prime }}v_{R^{\prime }}\left( r_{R^{\prime }}\right)
\sum_\Lambda \psi _{R^{\prime }\Lambda ,RL}^a\left( \varepsilon ,{\bf r}%
_R\right) .$

Returning to the strongly-scattering channels of $\sum_Rv_R\left( r_R\right)
\left[ \varphi _R^a\left( \varepsilon _i,{\bf r}_R\right) -\psi _i\left( 
{\bf r}\right) \right] ,$ if the overlap is so large that the $s_R$-sphere
overlaps a {\em neighboring} $a_{R^{\prime }L^{\prime }}$-sphere, then it is
simplest to imagine that we have {\em not} truncated the SSWs inside their
hard spheres, because otherwise the cancellation (\ref{canc}) would not take
place inside the $s_R$-$a_{R^{\prime }L^{\prime }}$ overlap. For consistency
then, we should not truncate the free partial waves $\varphi $ inside their
own hard sphere either. The resulting divergencies at the sites, of the SSWs
and of the free partial waves, of course cancel for the {\em smooth} linear
combinations. This undoing of the truncation inside the hard spheres is not
necessary, but it simplifies the bookkeeping.

The result (\ref{ovl1}) is then in agreement with what we found in Fig. \ref
{OVL}, that the error is a function which vanishes outside the regions of
overlap and that inside such a region, it is the product of a function, $%
v_{R^{\prime }}\left( r_{R^{\prime }}\right) ,$ which vanishes with a small
discontinuity at one of the MT spheres and a function, $\phi _R-\varphi _R,$
which vanishes quadratically at the surface of the other MT sphere, with a
prefactor proportional to the discontinuity of that MT potential. Remember
that the radial part of $\varphi _R$ is supposed to continue to the origin.
The result, which is given in (\ref{ovl2}), may be obtained from the radial
Schr\"{o}dinger and wave equations. Finally, in expression (\ref{ovl3}) we
have kept only the term of leading order and have used that $\phi _R^a\left(
\varepsilon _i,{\bf r}_R\right) \approx \Psi _i\left( {\bf r}\right) $ in
the region picked out by the other factors. Hence, {\em the error of the
wave function is of second order in the potential overlap.}

The error of the one-electron energy may be obtained by first order
perturbation theory as: $\Delta \varepsilon _i\equiv \varepsilon
_i-\varepsilon _i^{{\rm true}}\approx $ $-\left\langle \Psi _i\left| -\Delta
+\sum_Rv_R\left( r_R\right) -\varepsilon _i\right| \Psi _i\right\rangle $
and, to leading order, we find from Eq. (\ref{ovl3}) that the error of the
band-structure energy is\cite{Boston} 
\begin{eqnarray}
\sum_i^{{\rm occ}}\Delta \varepsilon _i &\sim &-\frac \pi
{24}\sum_{RR^{\prime }}^{{\rm pairs}}\left| {\bf R-R}^{\prime }\right|
^5\omega _{RR^{\prime }}^4\,v_R\left( s_R\right) v_{R^{\prime }}\left(
s_{R^{\prime }}\right) \rho \left( \frac{{\bf R+R}^{\prime }}2\right) ,
\label{ovl4} \\
{\rm where\quad }\omega _{RR^{\prime }} &\equiv &\frac{s_R+s_{R^{\prime }}}{%
\left| {\bf R-R}^{\prime }\right| }-1\quad {\rm is\;the\;radial\;overlap.}
\label{omega}
\end{eqnarray}
In a last section we shall demonstrate how this works for the
third-generation LMTO method. Finally, it may be noted that the appearance
of the KPW in Fig. \ref{KPW} is hardly influenced by the MT overlap. This
figure in fact applies to an overlap of $\omega {\rm =}14\%.$

Like the slope matrix, the kink matrix is not Hermitian, but the matrix 
\begin{equation}
K_{R^{\prime }L^{\prime },RL}^a\left( \varepsilon \right) \equiv
a_{R^{\prime }L^{\prime }}\left[ S_{R^{\prime }L^{\prime },RL}^a\left(
\varepsilon \right) -D\left\{ \varphi _{Rl}\left( \varepsilon ,a_{RL}\right)
\right\} \delta _{R^{\prime }L^{\prime },RL}\right]   \label{Kink}
\end{equation}
is.\cite{sign} This matrix is the renormalized screened KKR matrix. If we
multiply each of the kink-cancellation equations (\ref{canc}) with the
corresponding hard sphere radius, $a_{R^{\prime }L^{\prime }},$ these
equations take the form: $\sum K{\bf c}={\bf 0}$ and, hence, they are the
screened KKR equations. Just as the first energy derivative of the structure
matrix is the overlap matrix for the set of SSWs, so the first energy
derivative of the KKR matrix is the overlap matrix for the set of KPWs.\cite
{pm} In fact, one may show that the KKR matrix itself is the energy minus
the MT Hamiltonian in the basis of the KPWs with the same energy, that is, 
\begin{equation}
K_{R^{\prime }L^{\prime },RL}^a\left( \varepsilon \right) =\left\langle \Phi
_{R^{\prime }L^{\prime }}^a\left( \varepsilon \right) \left| \varepsilon
-\left( -\Delta +V\right) \right| \Phi _{RL}^a\left( \varepsilon \right)
\right\rangle .  \label{Ke}
\end{equation}

For Green-function and CPA calculations it has been very important that the
transformation (\ref{scale1}) of the resolvent, $\left[ {\sf P}\left(
z\right) -{\sf S}\right] ^{-1},$ from one representation to another is
merely a scaling rather than a matrix operation. This turns out to hold also
in the new formalism, and it means that such calculations may now be
performed with more realistic potentials and including downfolding. The
result is: 
\begin{equation}
K^b\left( z\right) ^{-1}=a^{-1}g^a\left( z,b\right) \varphi ^a\left(
z,b\right) +\varphi ^a\left( z,b\right) \,K^a\left( z\right) ^{-1}\,\varphi
^a\left( z,b\right)   \label{scal2}
\end{equation}
and has been obtained by use of the completeness relation (\ref{complete}),
the one-centre expansion (\ref{local}), and the following Wronskian
relations:\cite{Trieste} 
\[
ag^b\left( a\right) =-bg^a\left( b\right) ,\quad af^b\left( a\right)
=b^2g^a\left( b\right) ^{\prime },\quad a^2g^b\left( a\right) ^{\prime
}=bf^a\left( b\right) ,\quad a^2f^b\left( a\right) ^{\prime }=-b^2f^a\left(
b\right) ^{\prime },
\]
where the common energy argument, $z,$ has been dropped.

\bigskip

\noindent LOW-ENERGY, FEW-ORBITAL, TB HAMILTONIANS; HTSCs

\smallskip

If the energy dependence of the renormalized screened KKR matrix is
linearized around some chosen energy $\varepsilon _\nu ,$ 
\begin{equation}
K^a\left( \varepsilon \right) \approx \,K^a+\left( \varepsilon -\varepsilon
_\nu \right) \dot{K}^a=-\left\langle \Phi ^a\left| -\Delta +V\right| \Phi
^a\right\rangle +\varepsilon \left\langle \Phi ^a\mid \Phi ^a\right\rangle ,
\label{Klinear}
\end{equation}
then the KKR equations (\ref{KKR}) have the form of an algebraic eigenvalue
problem. In (\ref{Klinear}) and in the following, omission of an energy
argument $\varepsilon $ means that the function is evaluated {\em at} $%
\varepsilon _\nu .$ The basis set which, by use of the Raleigh-Ritz
variational principle for the MT Hamiltonian, gives rise to this problem
turns out to be the KPW-set at the fixed energy, $\varepsilon _\nu .$ This
follows from Eq. (\ref{Ke}) and is expressed in the second part of Eq. (\ref
{Klinear}). Since the off-diagonal elements of the overlap matrix, $\dot{K},$
only influence the energy eigenvalues to order $\left( \varepsilon
_i-\varepsilon _\nu \right) ^2,$ we may even neglect the non-orthogonality
of the KPWs and, for a crystal, obtain the correct Fermi surface, $%
\varepsilon _i\left( {\bf k}\right) {\rm =}\varepsilon _F
\equiv \varepsilon _{\nu },$ and the correct
group velocities, $\left. \partial \varepsilon _i\left( {\bf k}\right)
/\partial {\bf k}\right| _{\varepsilon _F}{\bf ,}$ by diagonalization of a
first-order Hamiltonian whose matrix elements are simply: $-\tilde{K}%
_{RL,R^{\prime }L^{\prime }}^a \equiv
\,-K_{RL,R^{\prime }L^{\prime }}^a \left/ \sqrt{%
\dot{K}_{RL,RL}^a \dot{K}_{R^{\prime }L^{\prime
},R^{\prime }L^{\prime }}^a }\right. .$ This
Hamiltonian is completely analogous to $h^{{\sf a}}$ in the ASA, but $-%
\tilde{K}^a$ implicitly contains the integrals over the interstitial region
and the downfolded channels, and it works to leading order in the overlap of
the potential wells. The range of $-K_{RL,R^{\prime }L^{\prime }}$ in $R$%
-space, and the size of the energy window inside which the linear
approximation holds, depends on the screening. Crudely speaking, the more
strongly-scattering channels included, and the larger their hard spheres
chosen without being touching, the shorter is the range of the hopping, and
the wider is the energy window.

$-\tilde{K}^a$ can be used as the low-energy, few-orbital, single-particle
part of correlated Hubbard-type Hamiltonians, as we shall now demonstrate
for a generic high-temperature superconductor (HTSC). We have in the past%
\cite{HTSC} been able to derive such a Hamiltonian for YBa$_2$Cu$_3$O$_7$
using the second-generation LMTO package.\cite{LMTO47} That procedure,
however, required a lot of hand-work and much insight, and has proved
cumbersome to use in general. The new procedure is far more automatic and
accurate,\cite{Tan} and has already proved successful for the ladder
compounds.\cite{Ladder}

The basic structural element of all HTSCs is a CuO$_2$ layer, which is a
quadratic lattice with copper at the corners and oxygen halfway between all
copper nearest neighbors. In the left-hand side of Fig. \ref{CaCuOorb}, the
copper sites are those which carry either a $d_{x^2-y^2}$ or an $s$ orbital,
and the oxygen sites are those which carry a $p_x$ or a $p_y$ orbital.
Different HTSC materials have different stackings of the CuO$_2$ layers with
various ''insulating'' and/or ''doping'' layers between them. Nevertheless,
the calculated LDA band structures near what is believed to be the Fermi
level of optimally doped HTSCs are very similar, and similar to that
calculated for the simplest possible such material; dimpled CaCuO$_2.$ In
this compound, the CuO$_2$ layers are stacked in the $z$-direction and are
separated by calcium, which sits in the hollow between the eight coppers of
the two neighboring layers. The oxygens in the Cu rows running in the $x$- ($%
y$-)direction are dimpled out of the plane by $+\left( -\right) \,7$
degrees. The right-hand side of Fig. \ref{CaCuOorb} shows a central CuO$_2$
layer seen from the side, with a $d_{x^2-y^2}$ and an $s$ orbital on the
copper sites and a $p_z$ orbital on the oxygen site. On the CuO$_2$ layer
above is shown a Cu $s$ orbital and on the CuO$_2$ layer below, an O $p_z$
orbital. Dimpled CaCuO$_2$ is a calculated structure,\cite{Savrasov} a
theorists dream which hardly exists in this simple form in nature. Its LDA
energy bands, which we shall now consider, are nevertheless very similar to
those calculated\cite{HTSC} for YBa$_2$Cu$_3$O$_7,$ one of the only known 
{\em stoichiometric} optimally doped HTSCs.

\input{infigCaCuOorb.tek}

At the Fermi level there is only {\em one} band per CuO$_2$ layer, and this
is the anti-bonding $pd\sigma $ band formed from the O$\,p_x$ -- Cu$%
\,d_{x^2-y^2}$ -- O$\,p_y$ orbitals. This band is at the top of the 10 eV
broad O$\,p$ -- Cu$\,d$ complex consisting of 16 bands, the upper (anti- and
non-bonding) part of which may be seen in Fig. \ref{CaCuObands} (a).
According to the LDA and the so-called Van Hove scenario of HTSC, the Fermi
level (zero in the figure) for the optimally doped compounds is very close
to the saddle-point of the conduction band at $\left( ak_x,ak_y\right)
=\left( \pi ,0\right) .$ Hybridization with the Cu $s$ band, which is 5 eV
above, has pushed this saddle-point of the anti-bonding $pd\sigma $ band
down in energy, to a point where it just ''straddles off'' the top of the
anti-bonding $pd\pi $ bands. This makes the structure susceptible to
out-of-row movements of oxygen, because this will mix $\sigma $ and $\pi $
bands. In particular the stable structures of CaCuO$_2$ and YBa$_2$Cu$_3$O$_7
$ have oxygen dimpled seven degrees out of the layer, and this mixes O$\,p_z$
character into the conduction band in such a way that its saddle-point at $%
\left( \pi ,0\right) $ becomes ''extended'' that is, the dispersion towards $%
\left( 0,0\right) $ becomes proportional to $k^4,$ {\it i.e. }{\em flat,}
while in the perpendicular direction, towards $\left( \pi ,\pi \right) ,$ it
remains $k^2.$ The mixing pushes the corresponding $pd\pi $ band down in
energy by about half an eV and leaves the top of other $pd\pi $ bands about
an eV below the Fermi level. We thus see that the orbital character of the
conduction band, which is the only one we wish to describe, is quite mixed.

\input{infigCaCuObands.tek}

The converged LDA bands are showed in panel (a) of Fig. \ref{CaCuObands}.
For comparison, panels (b)-(d) show the bands calculated with various
''minimal'' LMTO sets, specifically, with only the six O $p$ orbitals (b),
with only the Cu $d_{x^2-y^2}$ orbital (c), and with the six O $p$ orbitals
plus the Cu $s$ and $d_{x^2-y^2}$ orbitals (d). These four calculations all
employ the full 3rd-generation LMTO formalism, to be described in the
following section, in which the Hamiltonian and overlap matrices, (\ref
{LMTOH}) and (\ref{LMTOO}), are given in terms of $K^a\left( \varepsilon
_F\right) $ and its first three energy derivatives. Panel (b) and (c)
demonstrate the power of downfolding in the 3rd-generation LMTO scheme: One
may for instance completely leave out the Cu $d_{x^2-y^2}$ LMTOs by
attaching that partial-wave character to the tails of the neighboring O $p$
LMTOs (b), or one may completely leave out the O $p$ LMTOs, keeping per cell
just the one Cu $d_{x^2-y^2}$ LMTO whose tail then incorporates the O $p,$
Cu $s,$ and other characters (c). As one can imagine, such {\em massive}
downfolding leads to {\em long} range of the LMTOs. As an example, the
Fourier transform of the conduction band shown in panel (c) is the
two-centre Hamiltonian in the representation of orthogonalized Cu$%
\,d_{x^2-y^2}$ LMTOs, where the cone-like feature of the band around $\left(
0,0\right) ,$ caused by near degeneracy of the Cu$\,d_{x^2-y^2}$ and O$\,p_x$
orbital energies, gives rise to very long range. This long-ranged,
single-band Hamiltonian, we have called (the single-particle part of) the
''physical'' low-energy Hamiltonian.\cite{HTSC}

What we shall be interested in here is a ''chemical'' Hamiltonian, which has
short range and whose TB parameters behave in a meaningful way when the
structure is deformed and when we proceed to similar materials. Which
orbitals such a chemical Hamiltonian should contain is then dictated by the
range of the corresponding $K^a\left( \varepsilon \right) $ matrix. If we
imagine a Taylor series like (\ref{Klinear}), it is conceivable that the
higher energy-derivative matrices have longer range. We therefore expect to
obtain the shortest range when the energy region of interest is so small
that we only need $\tilde{K}_{RL,R^{\prime }L^{\prime }}^a\left( \varepsilon
_F\right) $ as defined above. For dimpled CaCuO$_2,$ the chemical basis set
turns out to be the one used to generate the bands shown in panel (d). For
the same eight orbitals, we show in panel (e) the bands calculated by
diagonalization of the effective two-center Hamiltonian $-\tilde{K}\left(
\varepsilon _F\right) .$ We see that this approximation conserves the shape
of the conduction band in the relevant range of energy. All computations
illustrated so far were converged in ${\bf R}$-space. In fact, they were
performed in ${\bf k}$-space, which means that we started out using the
Ewald method to compute $B\left( \kappa ,{\bf k}\right) .$ When we now
Fourier transform $\tilde{K}_{RL,R^{\prime }L^{\prime }}^a\left( \varepsilon
_F,{\bf k}\right) ,$\ we find that the only non-negligible matrix elements
are those given by the orbital energies and two-center hopping integrals in
Fig. \ref{CaCuOorb}. Panel (f) of Fig. \ref{CaCuObands} shows the
corresponding TB energy bands. This orthogonal, two-center TB Hamiltonian,
is seen to reproduce the conduction band very well and to give a
satisfactory description of the neighboring bands. This TB Hamiltonian,
which we have generated almost automatically, could also have been
calculated without the Ewald scheme, by inversion of Eq. (\ref{scr}) in $%
{\bf R}$-space. A bit of trial and error is still needed in finding an
optimal choice of the hard-sphere radii. The ones we used are listed in the
figure caption.

\bigskip

\noindent LINEAR MUFFIN-TIN ORBITALS

\smallskip

The first-order Hamiltonian $-\tilde{K}$ does not suffice to describe the
energy spectrum over the 10-20 eV range spanned by the valence and lower
conduction bands of strongly bonded materials. Nor does inclusion of terms
beyond the linear in the Taylor series (\ref{Klinear}) help, because this
does not lead to an algebraic eigenvalue problem. What is needed, is a set
of energy independent orbitals which, in contrast to the set of KPWs at a
fixed energy, is complete to {\em linear} order in $\varepsilon -\varepsilon
_\nu .$

From a set of KPWs, we first define a set of energy dependent MTOs: 
\begin{equation}
\left| \chi \left( \varepsilon \right) \right\rangle \equiv \,\left| \Phi
\left( \varepsilon \right) \right\rangle -\left| \dot{\Phi}\right\rangle 
\dot{K}^{-1}K\left( \varepsilon \right)  \label{MTO}
\end{equation}
Here and in the following we often drop the common superscript $a,$ and
omission of an energy argument means that $\varepsilon {\rm =}\varepsilon
_\nu .$ Moreover, we have used the notation in which $\left| \chi \left(
\varepsilon \right) \right\rangle $ is a row vector with elements $\left|
\chi _{RL}\left( \varepsilon \right) \right\rangle \equiv \chi _{RL}\left(
\varepsilon ,{\bf r}_R\right) $ and $K$ is a matrix. $\dot{\Phi}_{RL}\left( 
{\bf r}_R\right) $ is the first energy derivative at $\varepsilon _\nu $ of
the KPW, $\Phi _{RL}\left( \varepsilon ,{\bf r}_R\right) ,$ defined in (\ref
{KinkPW}). Since the hard spheres are kept independent of energy, the
strongly-scattering channels of the energy-derivative functions $\dot{\Phi}$
vanish at {\em all} the hard spheres. The $\dot{\psi}$-part is sketched in
the bottom half of Fig. \ref{SSW} and the Si $p_{x+y+z}$ MTO at energy $%
\varepsilon _v,$ that is the LMTO, is shown together with the corresponding
KPW in the right-hand side of Fig. \ref{KPW}.

The superposition of $\dot{\Phi}$-functions added to the KPW in (\ref{MTO})
is such as to make the MTO {\em smooth.} That this is so is seen immediately
by forming the kink matrix for the MTO: $K\left( \varepsilon \right) -\dot{K}%
\dot{K}^{-1}K\left( \varepsilon \right) =0.$ Still, the set of MTOs remains 
{\em complete} with respect to the MT potential, because with $\varepsilon _i
$ being the energy and ${\bf c}_i$ a corresponding solution of the KKR
equations, $K\left( \varepsilon _i\right) {\bf c}_i={\bf 0,}$ we find that
the same linear combination of MTOs is: $\left| \chi \left( \varepsilon
_i\right) \right\rangle {\bf c}_i=\,\left| \Phi \left( \varepsilon _i\right)
\right\rangle {\bf c}_i=\,\left| \Psi _i\right\rangle .$ In contrast to the
KPW, the MTO is {\em independent of energy to linear order }because by
differentiation of (\ref{MTO}) with respect to energy and subsequent setting 
$\varepsilon {\rm =}\varepsilon _\nu $ we get: $\left| \dot{\chi}%
\right\rangle =\left| \dot{\Phi}\right\rangle -\left| \dot{\Phi}%
\right\rangle \dot{K}^{-1}\dot{K}=0.$ The energy-independent set of LMTOs, $%
\left| \chi \right\rangle \equiv \,\left| \Phi \right\rangle -\left| \dot{%
\Phi}\right\rangle \dot{K}^{-1}K,$ is therefore complete to linear order
with respect to the MT Hamiltonian and therefore yields eigenvalues with
errors proportional to $\left( \varepsilon _i-\varepsilon _\nu \right) ^4.$
For comparison the conventional single-$\kappa $ LMTO set is complete to
zeroth order in the MT interstitial, albeit to first order in the spheres,
and therefore yields eigenvalue errors of order $\left( \varepsilon
_i-\varepsilon _\nu \right) ^2$ which originate from the interstitial. This
is illustrated in Fig. \ref{FE}. A price for carrying not only $\psi ,$ but
also $\dot{\psi}$ functions, is that the new LMTO sets corresponding to
different hard-sphere radii are no longer linear combinations of each other;
the wave-function error, $A^a\cdot \left( \varepsilon _i-\varepsilon _\nu
\right) ^2,$ has an $a$-dependent prefactor.

\input{infigFE.tek}

We now derive the expressions for the Hamiltonian and overlap matrices in
the new LMTO basis. For the integrals in all space of KPWs and their first
energy derivative functions, one obtains: $\left\langle \Phi \mid \Phi
\right\rangle =\dot{K},$ $\left\langle \Phi \mid \dot{\Phi}\right\rangle
=\left\langle \dot{\Phi}\mid \Phi \right\rangle =\frac 1{2!}\ddot{K},$ and $%
\left\langle \dot{\Phi}\mid \dot{\Phi}\right\rangle =\frac 1{3!}\stackrel{...%
}{K}.$ The LMTO{\em \ overlap matrix} is therefore: 
\begin{eqnarray}
\left\langle \chi \mid \chi \right\rangle  &=&\left\langle \Phi \mid \Phi
\right\rangle -\left\langle \Phi \mid \dot{\Phi}\right\rangle \dot{K}^{-1}K-K%
\dot{K}^{-1}\left\langle \dot{\Phi}\mid \Phi \right\rangle +K\dot{K}%
^{-1}\left\langle \dot{\Phi}\mid \dot{\Phi}\right\rangle \dot{K}^{-1}K 
\nonumber \\
&=&\dot{K}-\frac 1{2!}\left( \ddot{K}\dot{K}^{-1}K+K\dot{K}^{-1}\ddot{K}%
\right) +\frac 1{3!}K\dot{K}^{-1}\stackrel{...}{K}\dot{K}^{-1}K.
\label{LMTOO}
\end{eqnarray}
The matrix elements of the MT Hamiltonian used to generate the LMTO set may
be found in a similar way. Since the LMTO is smooth there are no problems
with Hermiticity like those occurring for the matrix elements between KPWs 
{\em alone}. What we mean is, that the result (\ref{Ke}) cannot be obtained
by naively taking matrix elements of an equation like: $\left[ H-\varepsilon
\right] \left| \Phi \left( \varepsilon \right) \right\rangle =0,$ where $%
H\equiv -\Delta +V,$ or of its energy derivative: $\left[ H-\varepsilon _\nu
\right] \left| \dot{\Phi}\right\rangle =\left| \Phi \right\rangle .$ For
matrix elements between {\em smooth} linear combinations of KPWs like: 
\begin{eqnarray}
\left\langle \chi \left| -\Delta +V-\varepsilon _\nu \right| \chi
\right\rangle  &=&\left\langle \Phi \left| H-\varepsilon _\nu \right| \Phi
\right\rangle -\left\langle \Phi \left| H-\varepsilon _\nu \right| \dot{\Phi}%
\right\rangle \dot{K}^{-1}K-K\dot{K}^{-1}\left\langle \dot{\Phi}\left|
H-\varepsilon _\nu \right| \Phi \right\rangle   \nonumber \\
&&+K\dot{K}^{-1}\left\langle \dot{\Phi}\left| H-\varepsilon _\nu \right| 
\dot{\Phi}\right\rangle \dot{K}^{-1}K  \nonumber \\
&=&-\left\langle \Phi \left| H-\varepsilon _\nu \right| \dot{\Phi}%
\right\rangle \dot{K}^{-1}K+K\dot{K}^{-1}\left\langle \dot{\Phi}\left|
H-\varepsilon _\nu \right| \dot{\Phi}\right\rangle \dot{K}^{-1}K  \nonumber
\\
&=&-\left\langle \Phi \mid \Phi \right\rangle \dot{K}^{-1}K+K\dot{K}%
^{-1}\left\langle \dot{\Phi}\mid \Phi \right\rangle \dot{K}^{-1}K  \nonumber
\\
&=&-K+\frac 1{2!}K\dot{K}^{-1}\ddot{K}\dot{K}^{-1}K  \label{LMTOH}
\end{eqnarray}
such procedures are however correct when used consistently for all terms.
Expression (\ref{LMTOH}) thus gives the {\em MT Hamiltonian matrix} which,
together with the overlap matrix (\ref{LMTOO}), are given exclusively in
terms of $K,\;\dot{K},\;\ddot{K},\,$and $\stackrel{...}{K}.$ These matrices
are square and labelled by the channels of the strong scatterers. We stress,
that in the 3rd-generation LMTO method, downfolding takes place at the
screening stage (\ref{scr}), where it removes the weakly-scattering channels
from the structure matrix $S\left( \varepsilon \right) .$ The calculations
for CaCuO$_2$ presented in Fig. \ref{CaCuObands} (a)-(d) employed this
formalism and convincingly demonstrated the new downfolding.

An approximation, which goes beyond the ASA and is not based on dividing
space into spheres and neglecting the remainder, consists of neglecting all
off-diagonal elements in the {\em real-}space representation of $\dot{K},\;%
\ddot{K},\,$and $\stackrel{...}{K.}$ With this {\em new} ASA, we have
avoided the matrix inversion, $\dot{K}^{-1},$ and the formalism contains
only {\em one} matrix, which we may take to be the first-order two-centre
Hamiltonian $-\tilde{K}$ defined in the previous section. This corresponds
to renormalizing each KPW and each MTO according to: $\left| \tilde{\Phi}%
_{RL}\left( \varepsilon \right) \right\rangle \equiv \,\left| \Phi
_{RL}\left( \varepsilon \right) \right\rangle \left/ \sqrt{\left\langle \Phi
_{RL}^2\right\rangle }\right. =\,\left| \Phi _{RL}\left( \varepsilon \right)
\right\rangle \left/ \sqrt{\dot{K}_{RL,RL}}\right. $ and $\left| \tilde{\chi}%
_{RL}\left( \varepsilon \right) \right\rangle \equiv \,\,\left| \chi
_{RL}\left( \varepsilon \right) \right\rangle \left/ \sqrt{\dot{K}_{RL,RL}}%
\right. ,$ and the rows and columns of the KKR matrix accordingly: $\tilde{K}%
_{RL,R^{\prime }L^{\prime }}\left( \varepsilon \right) \equiv
K_{RL,R^{\prime }L^{\prime }}\left( \varepsilon \right) \left/ \sqrt{\dot{K}%
_{RL,RL}\dot{K}_{R^{\prime }L^{\prime },R^{\prime }L^{\prime }}}\right. .$
With this renormalization, and taking $\varepsilon {\rm =}\varepsilon _\nu ,$
it is easy to see that expressions (\ref{LMTOO}) and (\ref{LMTOH}) reduce to
the simple ASA form (\ref{H1}) and (\ref{op1}).

We can develop an exact formalism by L\"{o}wdin orthonormalizing the KPWs,
instead of merely normalizing them: The overlap matrix for the renormalized
KPWs is: 
\begin{equation}
\left\langle \tilde{\Phi}_{RL}\mid \tilde{\Phi}_{R^{\prime }L^{\prime
}}\right\rangle =\stackrel{.}{\tilde{K}}_{RL,R^{\prime }L^{\prime }}\equiv
\delta _{RL,R^{\prime }L^{\prime }}+\Delta _{RL,R^{\prime }L^{\prime }},
\label{e6.30}
\end{equation}
where $\Delta $ is a Hermitian matrix with vanishing diagonal in $RL$-space.
Its off-site elements $\left( R\neq R^{\prime }\right) $ are usually
considerably smaller than unity and if we now define a Hermitian matrix: $%
\stackrel{.}{\tilde{K}}^{-1/2}=\,\left( 1+\Delta \right) ^{-\frac 12}\equiv
\,1-\frac 12\Delta +\frac 38\Delta ^2-...,$ which is the power-series
expansion in $\Delta ,$ then the linear combinations $\left| \bar{\Phi}%
\left( \varepsilon \right) \right\rangle \equiv \left| \tilde{\Phi}\left(
\varepsilon \right) \right\rangle \stackrel{.}{\tilde{K}}^{-1/2}$ are seen
to form an orthonormal set when $\varepsilon {\rm =}\varepsilon _\nu .$ This
is formally like in the conventional ASA. The partial waves truncated
outside and normalized inside the atomic $s$-spheres become in the formalism
of the 3rd generation the L\"{o}wdin orthonormalized kinked partial waves.
The transformed MTO set is: $\left| \bar{\chi}\left( \varepsilon \right)
\right\rangle \equiv \,\left| \tilde{\chi}_{RL}\left( \varepsilon \right)
\right\rangle \stackrel{.}{\tilde{K}}^{-1/2}=\,\left| \bar{\Phi}\left(
\varepsilon \right) \right\rangle +\left| \stackrel{.}{\bar{\Phi}}%
\right\rangle h\left( \varepsilon \right) ,$ where 
\begin{equation}
h\left( \varepsilon \right) \equiv -\stackrel{.}{\tilde{K}}^{-1/2}\tilde{K}%
\left( \varepsilon \right) \stackrel{.}{\tilde{K}}^{-1/2}=-\left( 1-\frac
12\Delta +\frac 38\Delta ^2-...\right) \tilde{K}\left( \varepsilon \right)
\left( 1-\frac 12\Delta +\frac 38\Delta ^2-...\right) .  \label{h}
\end{equation}
Since this expression for the MTO set is also formally identical with an
expression which, with the old definitions, was valid only in the ASA,
everything else works out the same. {\it E.g., }we find: $\left| \stackrel{.%
}{\bar{\chi}}\right\rangle =\,\left| \stackrel{.}{\bar{\Phi}}\right\rangle
+\left| \stackrel{.}{\bar{\Phi}}\right\rangle \dot{h}=0,$ because $\dot{h}%
=-1.$ The Hamiltonian and overlap matrices are thus given by (\ref{H1}) with 
$h\equiv h\left( \varepsilon _\nu \right) ,$ 
\begin{equation}
o\equiv \left\langle \bar{\Phi}\mid \stackrel{.}{\bar{\Phi}}\right\rangle
\;=\left\langle \stackrel{.}{\bar{\Phi}}\mid \bar{\Phi}\right\rangle =-\frac{%
\ddot{h}}{2!},\;{\rm and\;}p+o^2\equiv \frac 12\left\langle \bar{\Phi}\mid 
\stackrel{..}{\bar{\Phi}}\right\rangle =\frac 12\left\langle \stackrel{..}{%
\bar{\Phi}}\mid \bar{\Phi}\right\rangle =\left\langle \stackrel{.}{\bar{\Phi}%
}\mid \stackrel{.}{\bar{\Phi}}\right\rangle =-\frac{\stackrel{...}{h}}{3!}.
\label{op}
\end{equation}
In the 3rd-generation LMTO, $h,\;o,$ and $p$ are square matrices labelled by
the strongly-scattering channels. What we have accomplished is therefore to
transform the new Hamiltonian and overlap matrices, (\ref{LMTOH}) and (\ref
{LMTOO}), into the form (\ref{H1}), which was previously valid only in the
ASA.

In this language the {\em new} ASA corresponds to neglecting $\Delta $ as
well as the off-diagonal, real-space parts of $o$ and $p.$ A better
approximation is to keep $\Delta $ to first order in Eq. (\ref{h}), and {\em %
then} to neglect the off-diagonal parts in the real-space representation of $%
o$ and $p.$ In this way we still need to specify only {\em one} matrix,
namely the first-order, two-center TB Hamiltonian, $h,$ at the expense of
increasing its real-space range somewhat beyond that of $-\tilde{K}.$ In the
full formalism we have to specify 2 matrices, the Hamiltonian and the
overlap matrix or worse, the 3 matrices: $h,\;\left( \dot{h}=-1\right) ,\;%
\ddot{h},$ and $\stackrel{...}{h},$ or even worse, the 4 matrices: $K,\;\dot{%
K},\;\ddot{K},$ and $\stackrel{...}{K}$ whose real-space range increases
with the number of energy derivatives taken, that is, in order of decreasing
importance for the bands near $\varepsilon _\nu .$

\input{infigSi.tek}

Some of this is illustrated in Fig. \ref{Si} where we compare the LDA band
structure obtained from a converged 3rd-generation LMTO calculation (full
line) with results (dashed lines) obtained using various {\em minimal} basis
sets, $sp^3$ in (a)-(c) and $sp^3d^5$ in (d), and various {\em truncations.}
The empty-sphere $spd$- and, in (a)-(c), the Si $d$-channels were
downfolded. Here panel (a) demonstrates that it is possible with merely an $%
sp^3$ set to obtain an accurate first-principles description of the valence 
{\em and} four lowest conduction bands, provided that we allow the set to be
so long ranged that its Hamiltonian and overlap matrices, (\ref{LMTOH}) and (%
\ref{LMTOO}), extend to 12th-nearest neighbors. This basis is defined by: $%
a_s{\rm =}1.1t,\;a_p{\rm =}1.0t,$ and $\varepsilon _\nu {\rm =}-2$ eV. As
usual, $t$ is half the nearest-neighbor distance. If an accurate $sp^3$
TB-description is needed of merely the valence band, then it is possible to
limit the range of the orbitals to the extent that the Hamiltonian and
overlap matrices can be truncated after the 6th-nearest neighbors. In (b)
this is achieved mainly by shifting $\varepsilon _\nu $ down to the middle
of the valence band. In (c) and (d) we have simplified the calculation of
the Hamiltonian and overlap matrices by evaluating (\ref{h}) to only first
order in $\Delta ,$ and by neglecting the off-diagonal elements in $R$-space
of $o$ and $p.$ As mentioned above, this also makes it necessary to tabulate
only {\em one} two-centre matrix, $h.$ (Note that the {\em screened}
two-centre matrices cannot be completely specified by Slater-Koster
two-centre integrals like (\ref{two}), because the {\em screened }KPWs and
LMTOs do {\em not} have pure angular-momentum character). Comparison of the
dashed lines in (b) and (c) shows that this simplification works for the
valence-band structure, but that the quality of the conduction band, which
was not aimed at here, has deteriorated. So far we have not been able with
our first-principles procedure to find parameters which will decrease the
range of the $sp^3$ first-order two-centre Hamiltonian, $h,$ below
6th-nearest neighbors. However, with an $sp^3d^5$ basis this is possible,
because then also the $d$-channels can be used for screening. This is
demonstrated in panel (d), where the $sp^3d^5$-set with the parameters $a_s%
{\rm =}a_p{\rm =}1.0t,\;a_d{\rm =}0.9t,$ and $\varepsilon _\nu {\rm =}-6$
eV, plus the above-mentioned simplification, yields an $h$ which can be
truncated after 3rd-nearest neighbors. The resulting valence band is good
and the conduction band very reasonable.

\input{infigPOT.tek}

In the past there have been several attempts to model the energy bands of Si
by a simple TB Hamiltonian and the need for TB total-energy representations
to provide inter-atomic forces for molecular-dynamics simulations has
renewed this interest. These attempts range from simple nearest-neighbor,
orthogonal parametrization of diamond structured Si with an $sp^3$ basis in
the 70's\cite{Chadi} to recent work with long-ranged non-orthogonal $sp^3d^5$
basis sets\cite{Cohen} with a hope to provide transferable parameters. All
these works relied on fittings of energy bands and total energies obtained
from first-principles calculations. Our method is free from such fitting
procedures and is purely deterministic. The recent work of McMahan and
Klepeis\cite{Mc} is more similar in spirit to ours, but being based on a
full-potential multiple-kappa LMTO calculation with the need for subsequent
contraction to a minimal $sp^3d^5$ basis set, it is more complicated and
computationally far more demanding. In fact, our method is so fast, that for
us, transferability is no issue. But in all fairness, our total-energy and
force calculation is still pending.

\bigskip

\noindent GETTING RID OF THE EMPTY SPHERES

\smallskip

The full LDA potential for diamond-structured Si is shown in the top left of
Fig. \ref{POT}. What was used in the LMTO calculations of Fig. \ref{Si},
however, was the conventional ASA potential shown in the top-right panel of
Fig. \ref{POT}, which is slightly overlapping [$\omega $=14\%; see Eq. (\ref
{omega})] and, in addition to the Si-wells, has repulsive wells at the
E-sites to describe the hills of the potential. Despite its crude
appearance, this ASA SiE-potential, gives nearly exact LDA valence and
conduction bands. But this is a special case. In general, the potential
consists of spherically symmetric craters with hills in between, and the
latter can be of any shape. Such a potential is naturally modelled by a
superposition of atom-centered spherically-symmetric wells, and since we
have proved in Eq. (\ref{ovl1}) that the KKR method can handle such a
potential, unless the overlap is too large. The questions are whether this
holds also for the new LMTO method, and whether the overlap allowed by these
methods is sufficiently large that the MT zero moves up close to the hill
tops and the wave functions tail properly off into the voids. Non-MT
perturbations would then be local and simple to include. Therefore we first
try to treat diamond-structured Si. Two appropriate potentials with
respectively 30\% and 60\% radial overlap are shown in the bottom panels of
Fig. \ref{POT}.

\input{infigPOTr.tek}

We thus want to fit the full potential, ${\sf V}\left( {\bf r}\right) ,$ to
a constant (the MT zero) plus a superposition of spherical wells: $\;{\sf V}%
\left( {\bf r}\right) \sim V_{mtz}+\sum v_R\left( r_R\right) \equiv V\left( 
{\bf r}\right) .$ If we decide on a least-squares fit, that is, minimization
of $\left[ {\sf V}-V\right] ^2,$ then variation of the functions $v_R\left(
r_R\right) $ leads to a set of coupled integral equations, one for each $R$
saying that the {\em spherical} average around site $R$ for radius $r_R$
should be the same for the sum of the MT wells as for the full potential
minus $V_{mtz}.$ Variation of $V_{mtz}$ leads to one equation saying that
the average of the MT and the full potential should be the same. These
equations are fairly simple to solve numerically, but they do not quite
express what we want, because a volume element in a region like a void,
where the electron has little chance of being, enters with the same weight
in the fitting as a volume element in say the bond region. What we really
want is a {\em pseudo }potential which, for a certain band, say the valence
band, minimizes the mean squared deviation of the one-electron energies, $%
{\rm Tr}\rho \left[ {\sf H}-H\right] ^2={\rm Tr}\rho \left[ {\sf V}-V\right]
^2,$ and this then brings in the electron density, $\rho \left( {\bf r}%
\right) ,$ as weighting function. This weighting presents little problem for
the $\delta V_{mtz}$-equation, which is merely: $\int \left[ {\sf V}%
-V\right] \rho d^3r=0,$ but it complicates the $\delta v_R\left( r_R\right) $%
-equations so much, that we decided on keeping $\rho $ in the $\delta V_{mtz}
$-equation only. Our MT pseudo potential\cite{Catia} thus pseudizes the
hills rather than the core regions.

Since at this stage, we merely want to see whether we can get rid of the
empty spheres in the diamond structure by comparing the valence-band
structure calculated for the true potential with that calculated for its
pseudo potential, we take the true potential to be one for which we can
solve Schr\"{o}dinger's equation with high accuracy, namely the ASA
potential shown in the upper right of Fig. \ref{POT}. Since this potential
is discontinuous at the surfaces of the Si and E spheres, its pseudo
potentials, shown at the bottom, are not only discontinuous at $s,$ but also
at the Si AS radius. The radial behaviors of the Si and E wells of the ASA
potential, as well as that of the Si pseudo potential with 40\% radial
overlap, are shown in Fig. \ref{POTr}. By comparison of the pseudo
potentials with 30\% and 60\% radial overlap shown at the bottom of Fig. \ref
{POT}, it is obvious that the latter resembles the true potential most
closely. Whereas the MT zero of the 14\% overlapping ASA potential is only
slightly above the bottom of the valence band, that of the 40\% overlapping
pseudo potential lies 6 eV higher, and that of the 60\% overlapping
potential is at the top of the valence band.

\input{infigER.tek}

We have now used the new LMTO method [Eq.s (\ref{LMTOO}) and (\ref{LMTOH})
with a Si $sp^3d^5$ LMTO set and the Si $f$-channels downfolded] to
calculate the energy bands for the valence-band pseudo potentials as a
function of the radial overlap $\omega .$ The rms and mean errors of the
calculated valence bands are shown in Fig. \ref{ER} by diamonds. Since for
increasing $\omega ,$ the potential has increasing range and, hence,
increasing freedom, the {\em rms} error initially falls, but it eventually
rises again as the kinetic-energy errors given by Eq. (\ref{ovl1}) and
proportional to $\omega ^4$ take over. The minimum rms error of 80 meV per
electron is reached at 30\% overlap. The {\em mean} error we had expected to
vanish for overlaps so small that the kinetic-energy errors are negligible,
because the pseudo potential was constructed such that $\int \left[ {\sf V}%
-V\right] \rho d^3r=0.$ Nevertheless, the computation yields a
''background'' mean error of --50 meV per electron. This is most likely due
to errors of second order in ${\sf V}-V$ caused by the unphysical
discontinuities at the E-spheres of the ASA potential. We expect this
background error to vanish and thereby the rms error to be reduced, when for 
${\sf V}$ we use the full potential in the top left panel of Fig. \ref{POT}.

Since the kinetic-energy error is negative, it represents an attraction
between overlapping atoms, and this might cause problems in
molecular-dynamics calculations. However, although this attraction increases
rapidly with overlap, it does decrease for decreasing inter-atomic distance
and fixed $s$-radii [see Eq. (\ref{ovl4})].

If radial overlaps in excess of $\sim $30 \% are needed, then the
kinetic-energy error must be corrected. We have tried two schemes, the
results of which are given in Fig. \ref{ER} by the stars and the triangles.
In the first scheme (stars) we have merely modified the pseudo potentials by
including in the $\delta V_{mtz}$-equation the kinetic-energy error to
leading order as given by Eq. (\ref{ovl4}), whereby this equation becomes: $%
\int \left[ {\sf V}-V\right] \rho d^3r=\sum_i\Delta \varepsilon _i.$ This
leads to a reduction of the overlap error, mainly through reduction of the
discontinuity $v\left( s\right) .$ This correction is very simple, but as
seen from the figure, hardly sufficient because it only treats the error
proportional to $v\left( s\right) ^2\omega ^4.$ Our work on the second
scheme (triangles) is still in progress.\cite{Catia} Here, we evaluate the
LMTO Hamiltonian matrix properly to all orders in the overlap, that is, we
calculate the LMTO matrix elements following Eq. (\ref{ovl1}). Of course,
this adds terms to expression (\ref{LMTOH}) for the Hamiltonian and spoils
the beauty of Eq.s (\ref{H1}), (\ref{h}), and (\ref{op}), but we wish to
prove that we can control the overlap errors of the new LMTO method, and we
want to investigate how large overlaps we can handle. The preliminary
results shown in Fig. \ref{ER} are encouraging.

\end{document}

%% file: infigBoundary.tek
\begin{figure}[b!]
\unitlength1cm
\begin{minipage}[ht]{7.5cm}
\centerline{
\rotatebox{0}{\resizebox{2.5in}{!}{\includegraphics{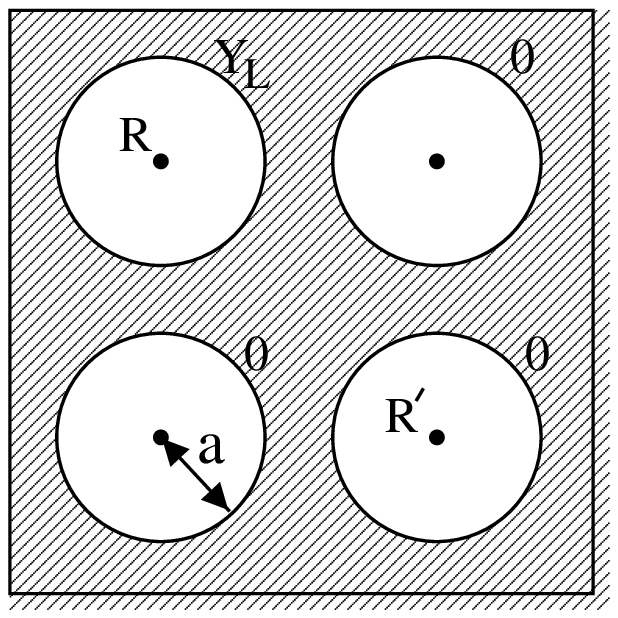}}}
}
\caption{\label{Boundary}
{\it (above)}. Boundary condition for the screened
spherical
wave, $\psi _{RL}^a\left( \varepsilon ,{\bf r}_R\right) .$ Only
strongly-scattering channels are indicated and all hard-sphere radii
are
equal.}

\caption{\label{SSW}
{\it (right).} Screened spherical wave centered at the
origin, $%
\psi _0^a\left( \varepsilon ,{\bf r}\right) ,$ and its slopes,
$S_{R,0}^a$ 
{\it (top).} The same for its first energy-derivative function,
$\dot{\psi}%
_0^a\left( \varepsilon ,{\bf r}\right) $ {\it (bottom)}.}

\end{minipage}
\hskip 1.0cm
\begin{minipage}[ht]{7.5cm}
\centerline{
\rotatebox{0}{\resizebox{2.5in}{!}{\includegraphics{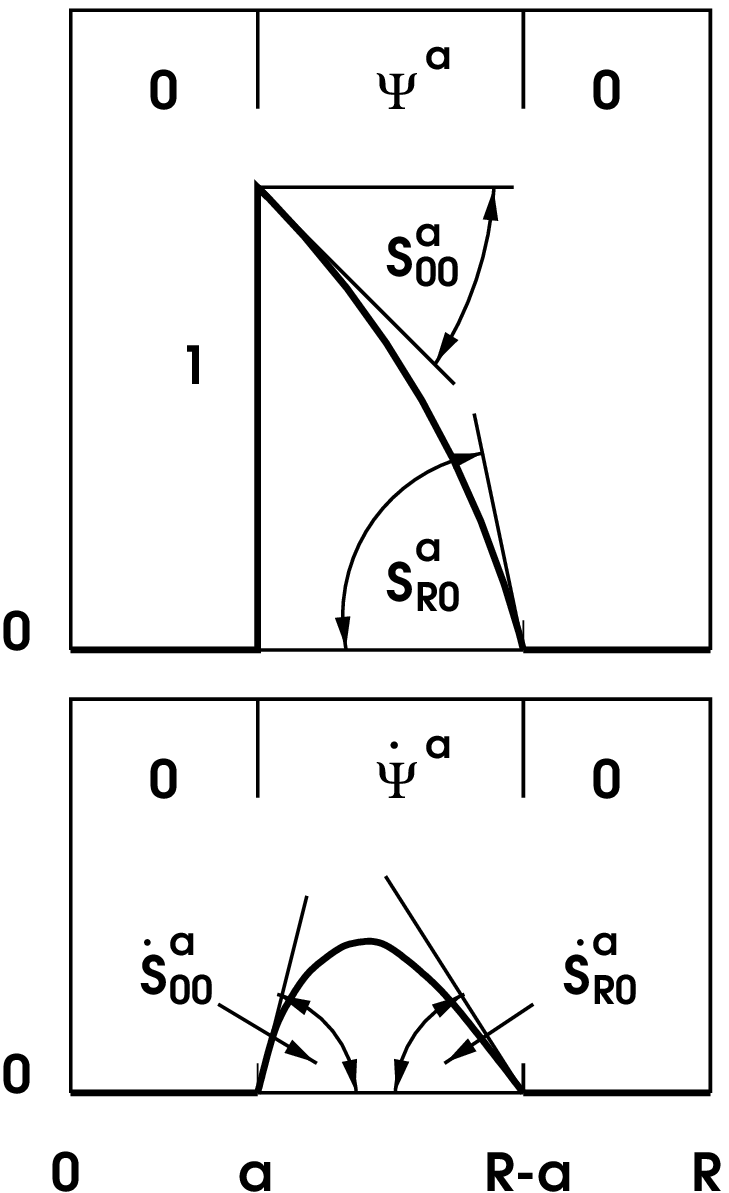}}}
}
\end{minipage}
\hfill
\end{figure}

%% file: infigS.tek
\begin{figure}[b!]
\unitlength1cm
\begin{minipage}[b]{16.cm}
\centerline{
\rotatebox{-90}{\resizebox{1.9in}{!}{\includegraphics{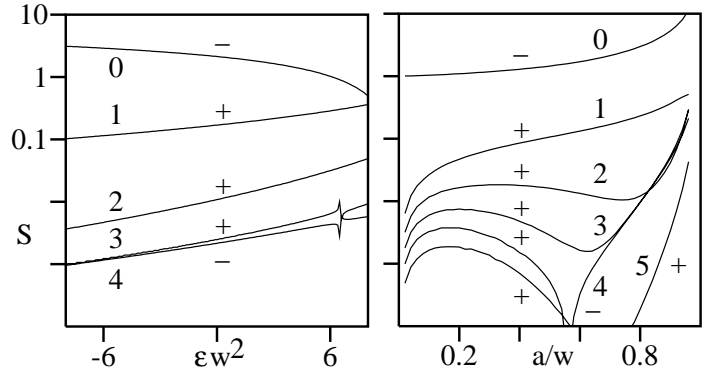}}}
}
\end{minipage}
\caption{\label{S}
The $ss\sigma $-element, $S_{R00,000}^a\left( \varepsilon
\right) ,
$ of the slope matrix for the fcc structure and $R$ in the 0'th to 4'th
or
5'th shell. {\it Left: }$a_{spd}$=0.7$w.$ Here $w$ is the Wigner-Seitz
radius, which in the fcc structure is 10\% larger than that of touching
spheres. $\varepsilon w^2$=6.05 corresponds to the lowest free-electron
energy at the $X$ point. {\it Right:} $\varepsilon $=0 and $a_{spd}{\rm
=}a.$
The $S$-scale is logarithmic and the channels with $l>2$ were taken as
non-scattering. The number of atoms in the 0'th--5'th shell are
respectively
1, 12, 6, 24, 12, and 24. The calculation was performed by matrix
inversion,
Eq. (\ref{scr}),
in real space for a 79-site cluster. For positive energies
it was necessary to prevent resonances at the surface of this cluster
by
enclosing it in a concave sphere simulating the boundary condition
$\psi $=0
on the spheres outside the cluster and carrying max$\left( l\right)
$=8. The
artefact at $\varepsilon w^2\approx $6.3 is a surviving resonance. The
results are accurate only when the SSW is well localized within the
cluster.
For cases where the SSW's are not localized within a cluster of
affordable
size, we must assume crystalline boundary conditions and Bloch-sum
$B\left(
\kappa \right) $ with the Ewald technique before performing the
screening
inversion (\ref{scr}) for each ${\bf k}$-point. With largely
overlapping MT
spheres, the energies of occupied states are always negative.}
\hfill
\end{figure}

%% file: infigOVL.tek
\begin{figure}[b!]
\unitlength1cm
\begin{minipage}[ht]{8.5cm}
\centerline{
\rotatebox{0}{\resizebox{3.3in}{!}{\includegraphics{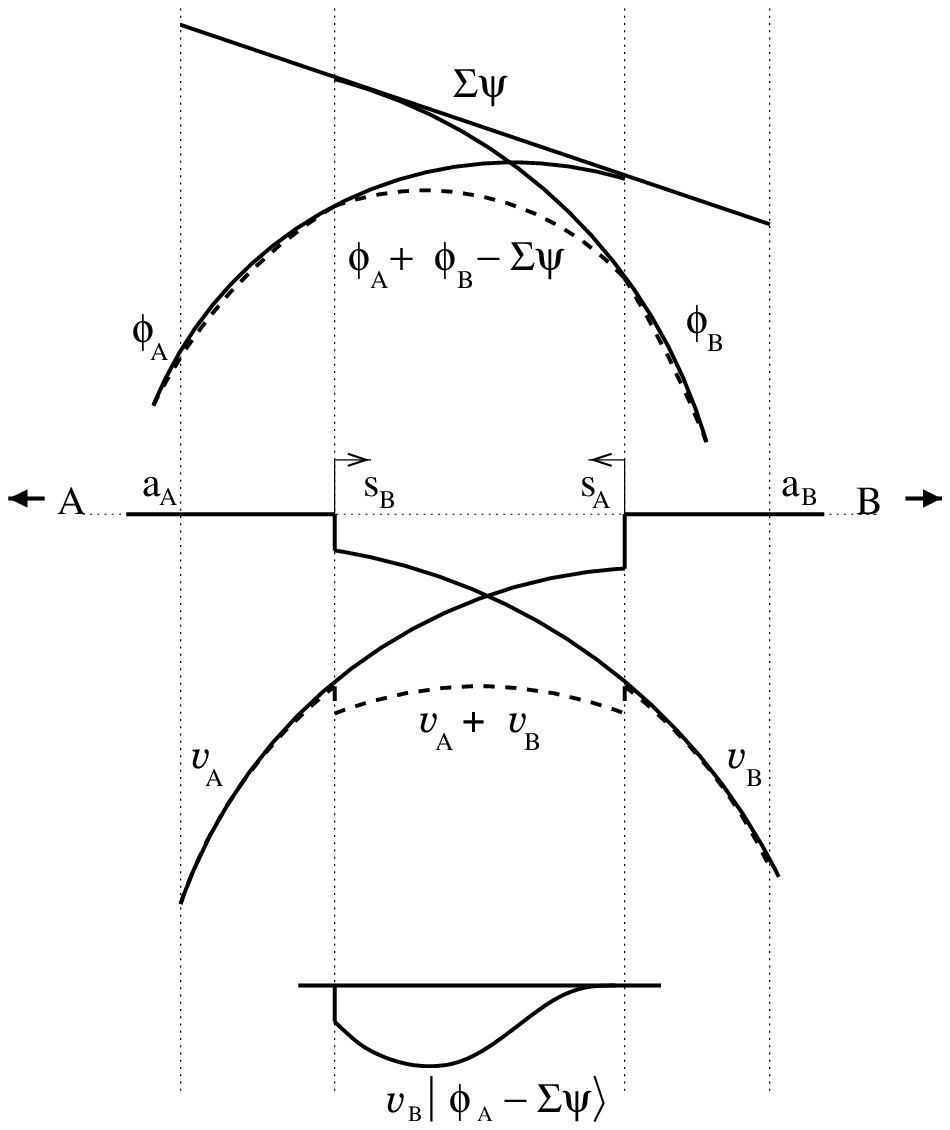}}}
}
\end{minipage}
\hskip 1.0cm
\begin{minipage}[ht]{6.9cm}
\caption{\label{OVL}
{\it Middle:} Overlapping potential wells, $v_A\left(
r_A\right)
$ and $v_B\left( r_B\right) ,$ centered at sites ${\bf A}$ (far left)
and $%
{\bf B}$ (far right). {\it Top:} The solution $\phi _A\equiv \sum_L\phi
_{AL}\left( \varepsilon _i,{\bf r}_A\right) c_{AL,i}$ joins smoothly
onto
the free solution $\varphi _A$ at $s_A.$ $\varphi _A$ runs backwards to
$a_A,
$ where the kink with the interstitial solution, $\sum \psi \equiv
\sum_{RL}\psi _{RL}\left( \varepsilon _i,{\bf r}_R\right) c_{RL,i},$ is
cancelled. Similarly for $\phi _B$ and $\varphi _B.$ Due to kink
cancellation, the resulting wave function, $\phi _A-\varphi _A+$ $\phi
_B-\varphi _B+\sum \psi ,$ equals $\phi _A+$ $\phi _B-\sum \psi $ in
this
picture where the angular-momentum character has been suppressed. {\it
Bottom: }The error, $\left[ -\Delta +v_A+v_B-\varepsilon _i\right]
\left|
\phi _A+\phi _B-\sum \psi \right\rangle$ \protect\newline
$=v_B\left| \phi _A\right\rangle
+v_A\left| \phi _B\right\rangle -\left( v_A+v_A\right) \left| \sum \psi
\right\rangle$ \protect\newline
$=v_B\left| \phi _A-\sum \psi \right\rangle +v_A\left|
\phi
_B-\sum \psi \right\rangle \,,$ consists of two terms each of which,
{\it 
e.g.} the first,\ is the product of $v_B\left( r_B\right) $ and $\phi
_A-\sum \psi ,$ which vanishes like $v_A\left( s_A\right) \left(
s_A-r_A\right) ^2\phi _A\left( s_A\right) $ at the $s_A$ boundary.
Hence
the error is of {\it second} order in the potential overlap.}
\end{minipage}
\hfill
\end{figure}

%% file: infigCaCuOorb.tek
\begin{figure}[b!]
\unitlength1cm
\begin{minipage}[ht]{16.cm}
\centerline{
\rotatebox{0}{\resizebox{4.6in}{!}{\includegraphics{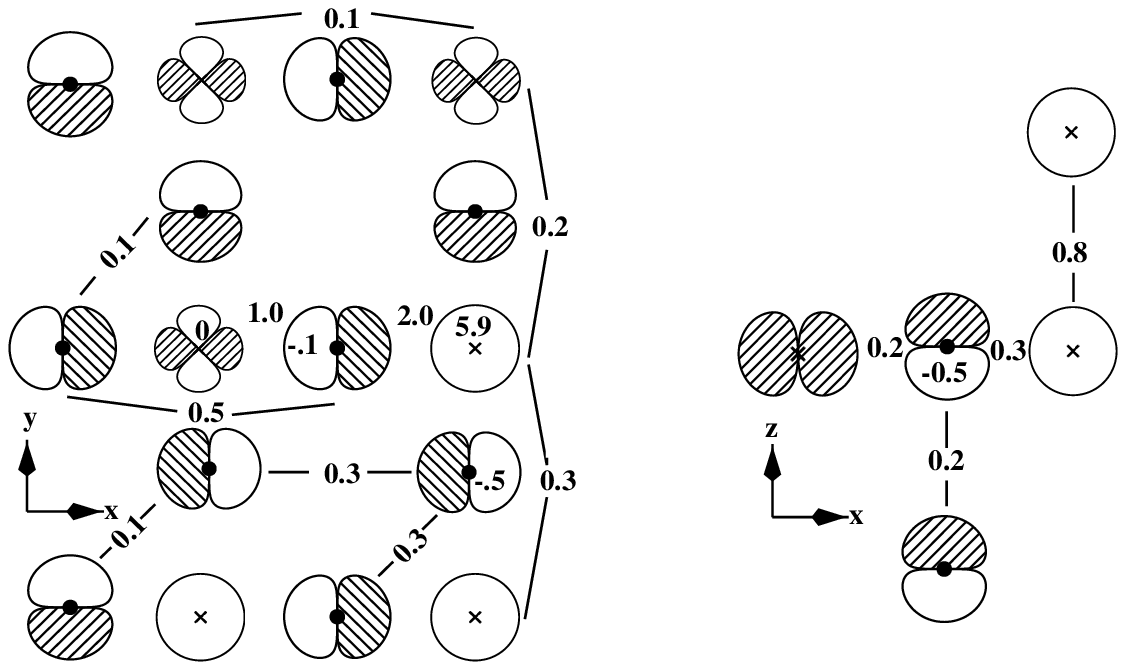}}}
}
\caption{\label{CaCuOorb}
The eight orbitals [O$_{{\rm x}}$ $p_{x,y,z},$
O$_{{\rm y}}$
$p_{x,y,z,},$ Cu $s,$ and Cu $d_{x^2-y^2}]$ and the values of their
energies
(with respect to $\varepsilon _{x^2-y^2})$ and two-centre hopping
integrals
(eV). These values were obtained as the matrix elements of $-\tilde{K}%
^a\left( \varepsilon _F\right) $ with all channels other than the eight
downfolded.}
\end{minipage}
\hfill
\end{figure}

%% file: infigCaCuObands.tek
\begin{figure}[b!]
\unitlength1cm
\begin{minipage}[ht]{16.cm}
\centerline{
\rotatebox{0}{\resizebox{5.5in}{!}{\includegraphics{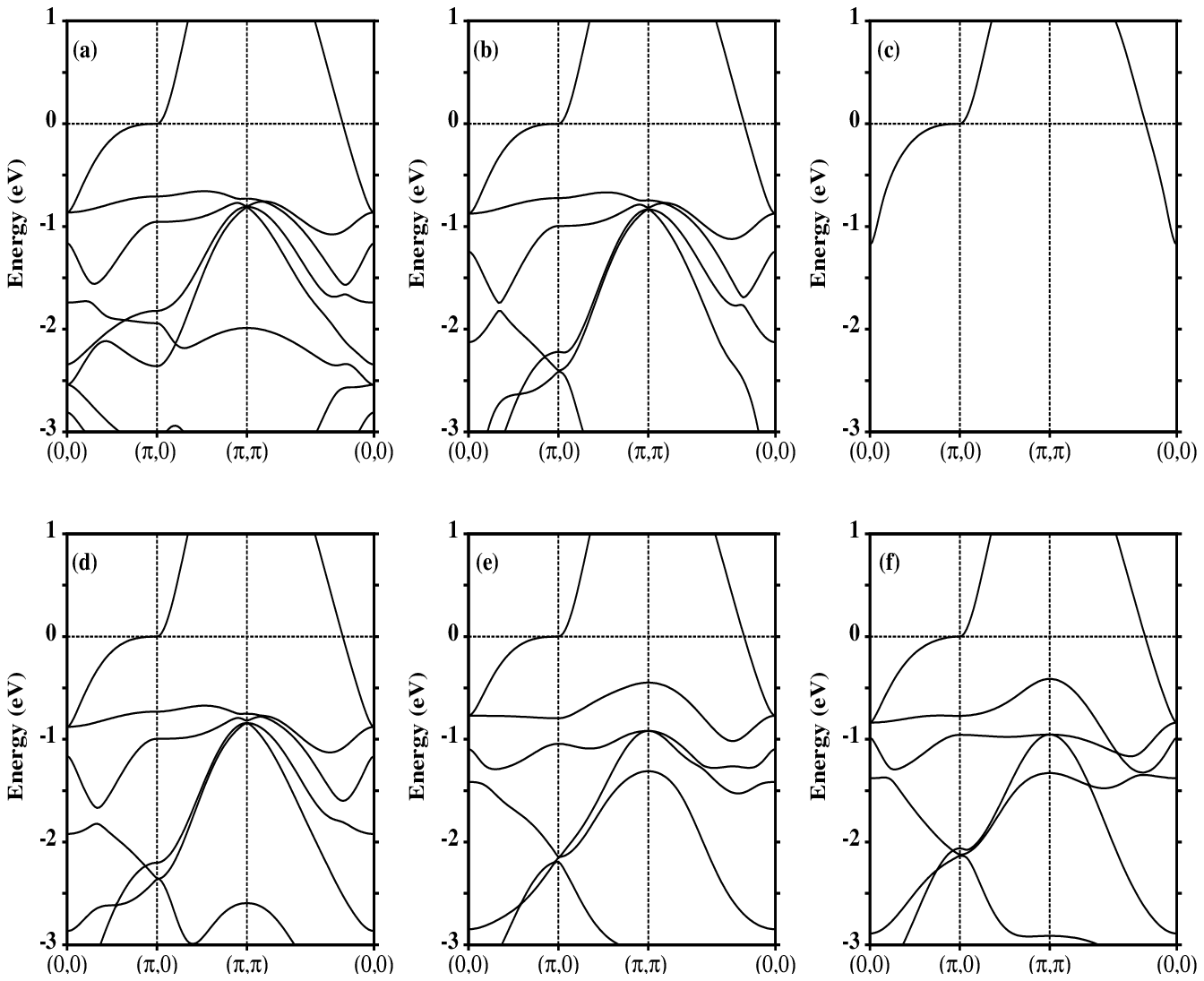}}}
}
\caption{\label{CaCuObands}
LDA energy bands for dimpled CaCuO$_2$ calculated
with
the 3rd-generation LMTO method using a converged LMTO basis, {\it (a),}
and
five different simplifications, {\it (b)-(f).} Reciprocal-space
distances
are in units of the reciprocal of the Cu-Cu distance. $\varepsilon
_F{\rm =0=%
}\varepsilon _\nu .$\ {\it (a):} All but the Cu$\,s$ and $d,$ and the
O$\,p$
orbitals downfolded. $a_{{\rm Cu\,}s}{\rm =}a_{{\rm Cu\,}d}{\rm
=}0.87t,\;a_{%
{\rm O\,}p}{\rm =}0.75t,$ where the touching-sphere radius, $t,$ is 1/4
the
Cu-Cu distance. {\it (b):} All but the six oxygen $p$ orbitals
downfolded. $%
a_{{\rm O\,}p}{\rm =}0.96t.$\ {\it (c):} All but the single
Cu$\,d_{x^2-y^2}$
orbital downfolded. $a_{{\rm Cu\,}x^2-y^2}{\rm =}0.62t.\;${\it (d):}
All but
the six oxygen orbitals, the Cu$\,s,$ and the Cu$\,d_{x^2-y^2}$
orbitals
downfolded. $a_{{\rm Cu\,}s}{\rm =}0.87t,\;a_{{\rm Cu\,}x^2-y^2}{\rm
=}%
0.68t,\;a_{{\rm O\,}p}{\rm =}0.96t,$ {\it (e):} Like {\it (d),} but
with $%
H-\varepsilon _F=-\tilde{K}$ and $O=1.$ {\it (f):}{\bf \ }Like {\it
(e),}
but with $\tilde{K}$ truncated after 3rd-nearest-neighbor hoppings,
that is,
with the orbital energies and two-centre hopping integrals given in
Fig. \ref
{CaCuOorb}.}
\end{minipage}
\hfill
\end{figure}

%% file: infigFE.tek
\begin{figure}[b!]
\unitlength1cm
\begin{minipage}[ht]{16.cm}
\centerline{
\rotatebox{-90}{\resizebox{2.8in}{!}{\includegraphics{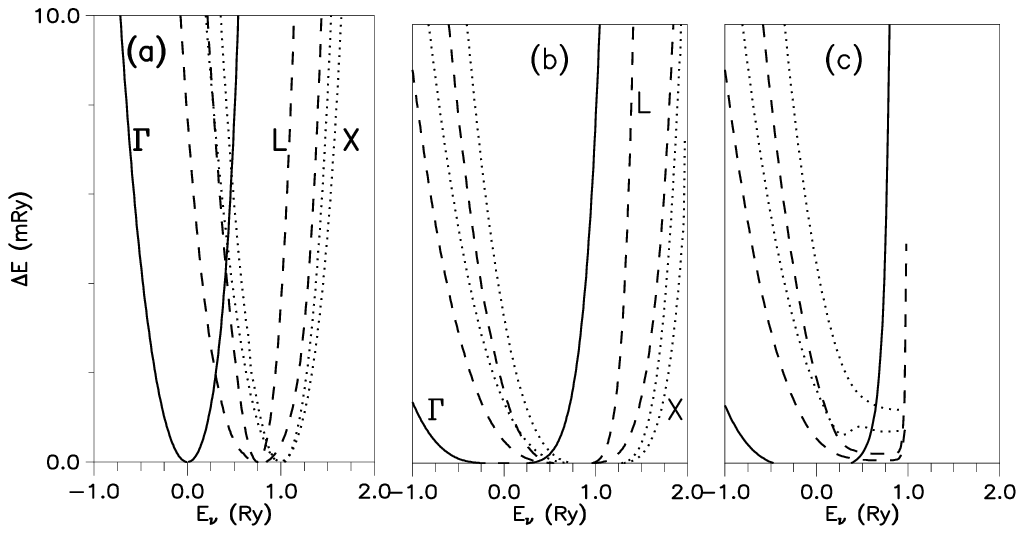}}}
}
\caption{\label{FE}
LMTO errors of the lowest free-electron energies (0,
0.75,
and 1 Ry) at the fcc $\Gamma ,\;$L, and X points as functions of the
energy-expansion parameter $\varepsilon _\nu $. The states at L and X
are
doubly degenerate and split when $\varepsilon _\nu \neq \varepsilon .$
{\it %
(a):} Old LMTO method with $spd$-basis and $s$=$w.$ {\it (b):\ }New
LMTO
method, Eq.s (\ref{LMTOO}) and (\ref{LMTOH}), with $spd$-basis and
$a_{spd}%
{\rm =}0.7w{\rm =}0.77t;$ the $s$-value is irrelevant. In {\it (a)} and
{\it %
(b)} the bare structure matrix was Bloch-summed with the Ewald
technique
before it was screened by matrix inversion, Eq. (\ref{scr}), in ${\bf
k}$%
-space. {\it (c):} Like {\it (b)} except that the inversion (\ref{scr})
was
performed in ${\bf R}$-space using a 79-site cluster enclosed in a
concave
sphere.}
\end{minipage}
\hfill
\end{figure}

%% file: infigSi.tek
\begin{figure}[b!]
\unitlength1cm
\begin{minipage}[ht]{7.5cm}
\centerline{
\rotatebox{-90}{\resizebox{2.9in}{!}{\includegraphics{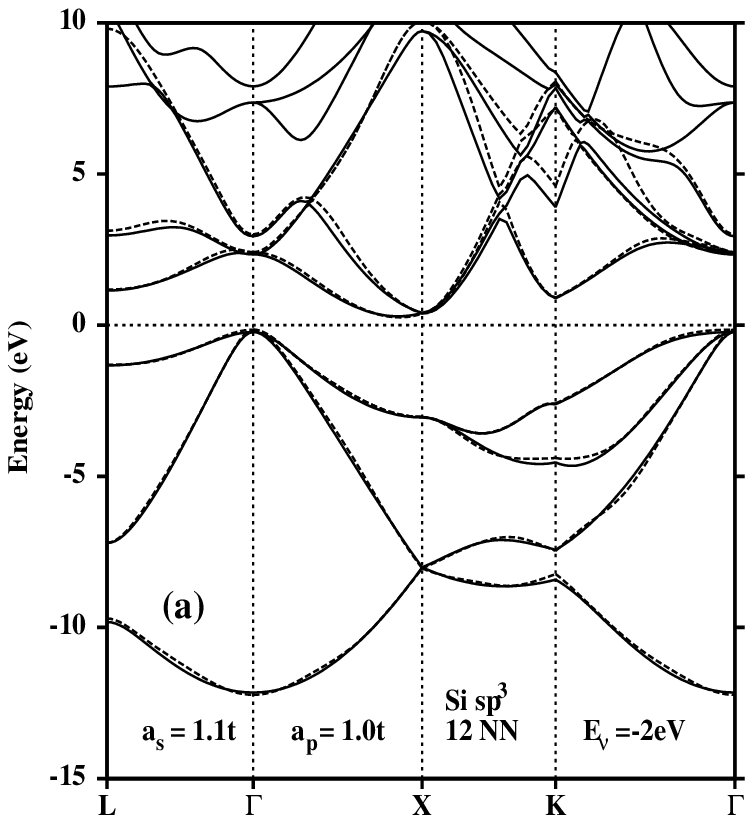}}}
}
\end{minipage}
\hskip 1.0cm
\begin{minipage}[ht]{7.5cm}
\centerline{
\rotatebox{-90}{\resizebox{2.9in}{!}{\includegraphics{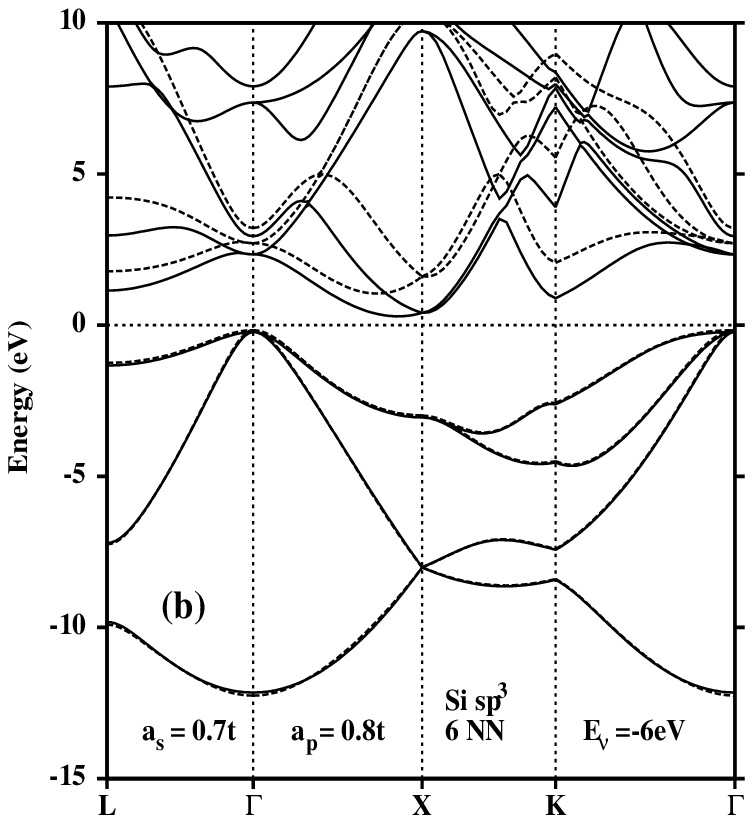}}}
}
\end{minipage}
          
\begin{minipage}[ht]{7.5cm}
\centerline{
\rotatebox{-90}{\resizebox{2.9in}{!}{\includegraphics{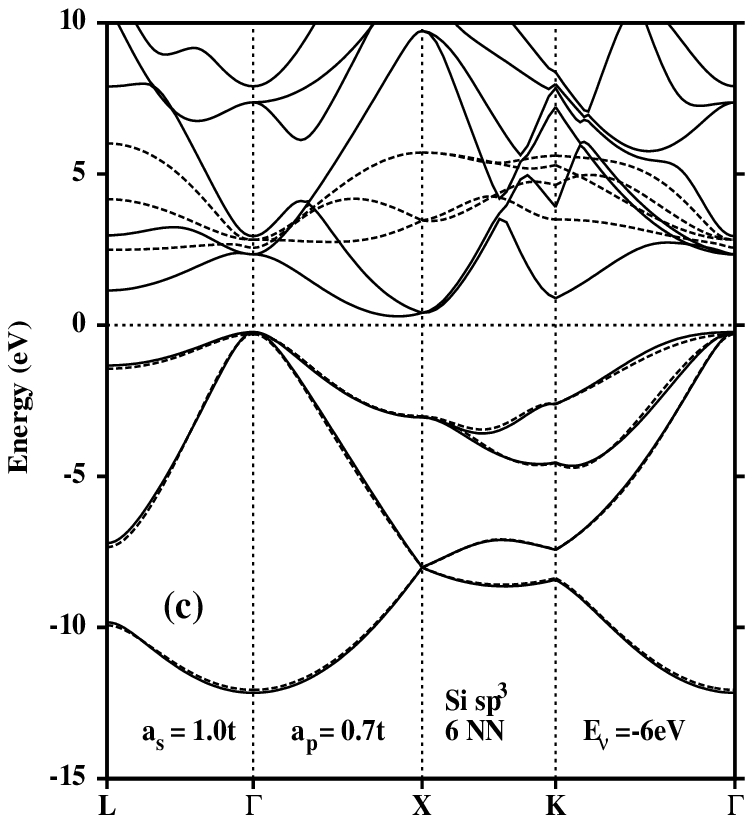}}}
}
\end{minipage}
\hskip 1.0cm
\begin{minipage}[ht]{7.5cm}
\centerline{
\rotatebox{-90}{\resizebox{2.9in}{!}{\includegraphics{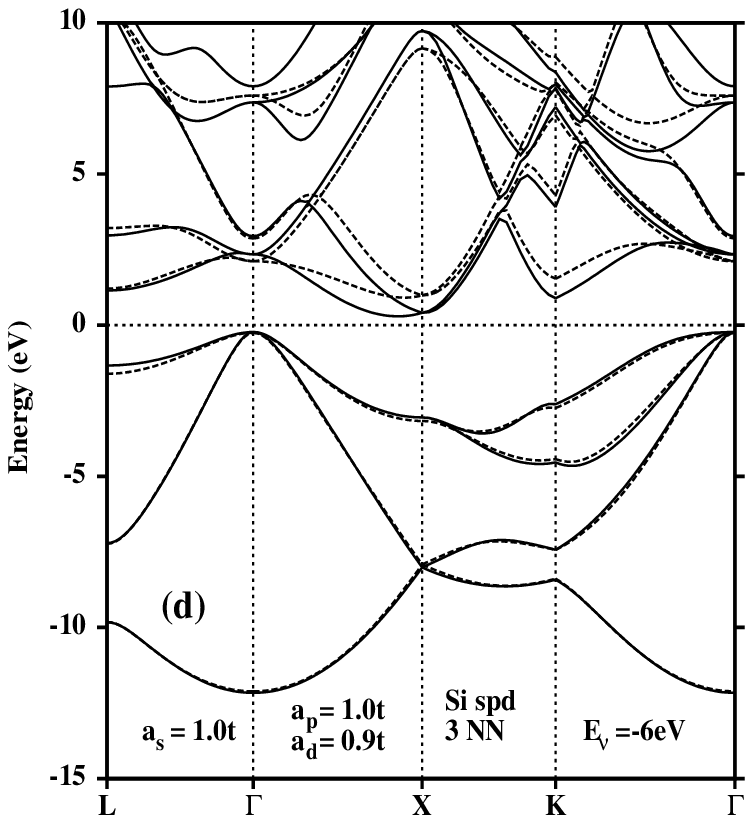}}}
}
\end{minipage}
\caption{\label{Si}
%{\footnotesize
LDA energy bands of diamond structured Si in full lines and
various TB approximations thereto in dashed lines (a)-(d). The
corresponding
LMTO sets and the real-space truncation of the Hamiltonian and overlap
matrices (\ref{LMTOH}) and (\ref{LMTOO}) in (a) and (b), and of $h$
(\ref{h}) in (c) and (d), are specified at the bottom of the panels.
}%}
\hfill
\end{figure}

%% file: infigPOTr.tek
\begin{figure}[b!]
\unitlength1cm
\begin{minipage}[ht]{8.5cm}
\centerline{
\rotatebox{0}{\resizebox{3.5in}{!}{\includegraphics{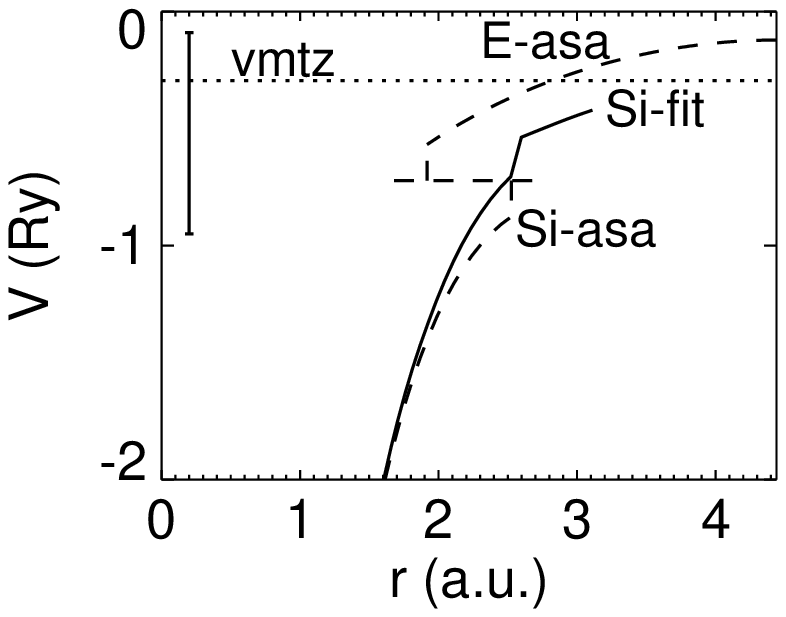}}}
}
\end{minipage}
\hskip 1.0cm
\begin{minipage}[ht]{6.9cm}
\caption{\label{POTr}
Radial behavior of the Si and E potential wells of the ASA
potential (dashed), and of the Si well of the pseudo potential with
40\%
radial overlap (full). The corresponding 3-dimensional potentials are
shown
in the upper right and lower left of Fig. \ref{POT}. The dotted line is
the
MT zero of the pseudo potential. The vertical bar at the left-hand side
indicates the position and extent of the valence band.}
\end{minipage}
\hfill
\end{figure}

%% file: infigER.tek
\begin{figure}[b!]
\unitlength1cm
\begin{minipage}[ht]{7.5cm}
\centerline{
\rotatebox{0}{\resizebox{2.7in}{!}{\includegraphics{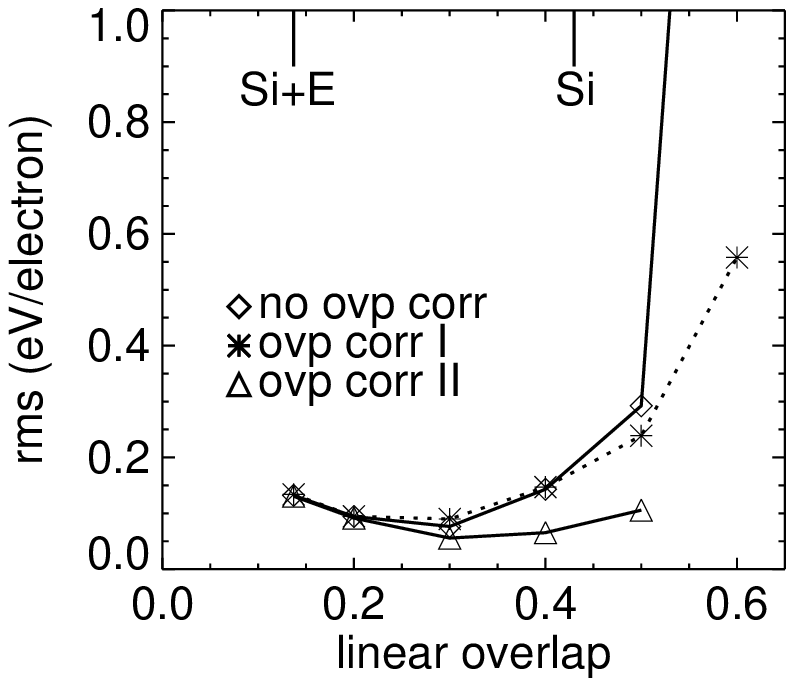}}}
}
\end{minipage}
\hskip 1.0cm
\begin{minipage}[ht]{7.5cm}
\centerline{
\rotatebox{0}{\resizebox{2.7in}{!}{\includegraphics{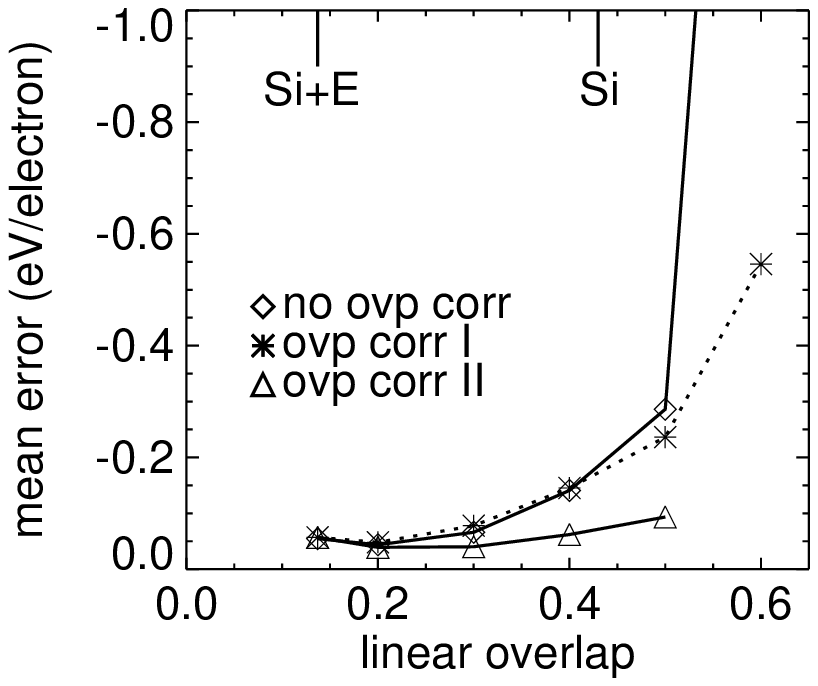}}}
}
\end{minipage}
\caption{\label{ER}
Rms and mean errors of the valence-band energies arising by
pseudizing of the ASA potential of diamond structured Si in order to
get rid
of the E wells. The abscissa is the radial overlap, $\omega ,$ as
defined in
Eq. (\ref{omega}). For {\em space-filling} Si+E spheres, $\omega {\rm
=}$%
14\%, and for Si spheres 43\%. Overlap correction I modifies the pseudo
potential, while II includes the proper kinetic energy in the LMTO
Hamiltonian.}
\hfill
\end{figure}

%% file: mrs977.bbl
\begin{thebibliography}{99}
\bibitem{sign}  In order to make $\dot{K}$ positive-, rather than negative
definite, we have defined $K$ with the opposite sign as in Ref. 3.
\end{thebibliography}
